\def\astroph{1}

\ifnum\astroph=0
\documentclass[12pt,preprint]{aastex}
\else
\documentclass{emulateapj}
\fi

\usepackage{natbib}
\usepackage{amsmath}
        \newcommand{\CAPI}{
                {Radiative cooling function (full) and
                implemented cooling function (dashed) with 
                quenched  cooling in the photospheric and chromospheric
                temperature regimes}
        }

        \newcommand{\CAPII}{
                {Velocity amplitude versus spherical harmonic
                wavenumber for the three layer granulation model (full)
                with the error bars showing maximum deviation through
                a period of 1 hour. The dashed line shows $u \propto
                k$.} 
        }

        \newcommand{\CAPIII}{
                {Three snapshots of 3000 corks released in the
                simulated velocity field, without magnetic velocity
                quenching, at $t=0$ min (left), $t=15$ min
                (middle), and $t=45$ min (right)}
        }

        \newcommand{\CAPIV}{
                {Initial stratification for density (full),
                gas pressure (dashed) and
                temperature (dotted). Also shown at the top is
                the vertical-scale, with a small line at each grid point}
        }

        \newcommand{\CAPV}{
                {Horizontally averaged energy deposition from resistive
                dissipation (full), Spitzer conductivity (long dash),
                convective flux (dash dotted) and radiative cooling
                (dotted), as a function of height} 
        }

        \newcommand{\CAPVII}{
                {PDF of the electric current
                squared as a function of height, dark colours mean
                higher PDF with logarithmic colour table. Over-plotted is
                the horizontal average of the magnetic field squared
                (dotted) and the horizontal average of the current
                squared (dashed)} 
        }

        \newcommand{\CAPVI}{
                {Images of the scalar product of electric current and 
                the magnetic field, for three
                different heights in the atmosphere, 0.0 \un{Mm},
                3.0\un{Mm} and 5.6\un{Mm} corresponding to
                photosphere, upper chromosphere and low corona (left
                to right). On top are two contours with values a factor
                of ten apart (black: high, white: low)}
        }

        \newcommand{\CAPIX}{ {Energy conversion efficiency as function of time
            during a 23 min period. The relative amount of energy converted into heating
            (solid) and the relative amount of energy lost to radiation
            (dashed) in the corona are shown} 
        }

        \newcommand{\CAPVIII}{ {Dissipated energy as function of time scale in
            the upper chromosphere (full), transition region (long dashes),
            lower corona (dashes) and upper corona (dotted) in arbitrary
            units. The curves are normalised at the maximum time
            scale} 
        }

        \newcommand{\CAPX}{ {Average heating rate in UV to X-ray emitting gas
            as a function of the underlying photospheric magnetic field
            strength threshold} 
        }

        \newcommand{\CAPXI}{ {Surface showing the height of the transition
            region, defined as where the temperature rises above $10^5
            \un{K}$}  
        }

        \newcommand{\CAPXII}{ {Histogram of the temperature PDF for each
            height for the FAL--C chromosphere. Dark is high PDF, with a
            logarithmic grey scale. The horizontal average is over plotted
            (dashed)} 
        }

        \newcommand{\CAPXIII}{ {Histogram of the density PDF for each height
            for the FAL--C chromosphere. Dark is high PDF, with a logarithmic
            grey scale. The horizontal average for the distribution is over plotted
            (dashed)} 
        }

        \newcommand{\CAPXIV}{ {Emulated TRACE 171 (left top) and 195 (right
            top) images both spanning 1.5 box lengths, and the underlying
            photospheric magnetic field} 
        }

\ifnum\astroph=0

        \newcommand{\FIGI}{
                \begin{figure}[t]
                \figurenum{1}
                \plotone{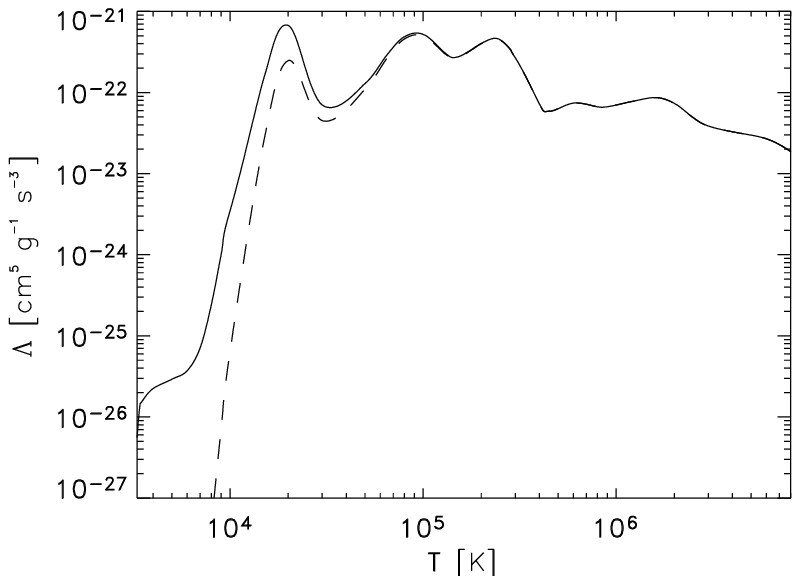}
                \caption{\CAPI}
                \label{fig:lambda}
                \end{figure}
        }

        \newcommand{\FIGII}{
                \begin{figure}[t]
                \figurenum{2}
                \plotone{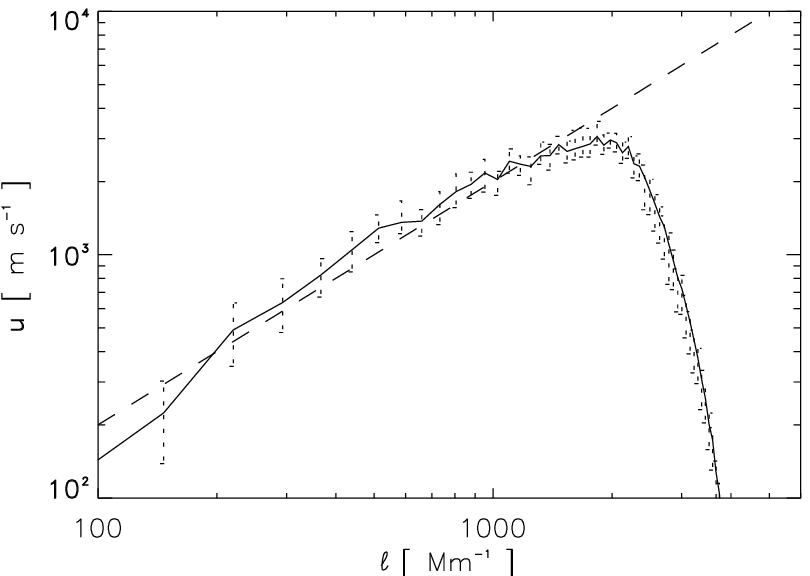}
                \caption{\CAPII} 
                \label{fig:vspectrum}
                \end{figure}
        }

        \newcommand{\FIGIII}{
                \begin{figure}[t]
                \figurenum{3}
                \plotone{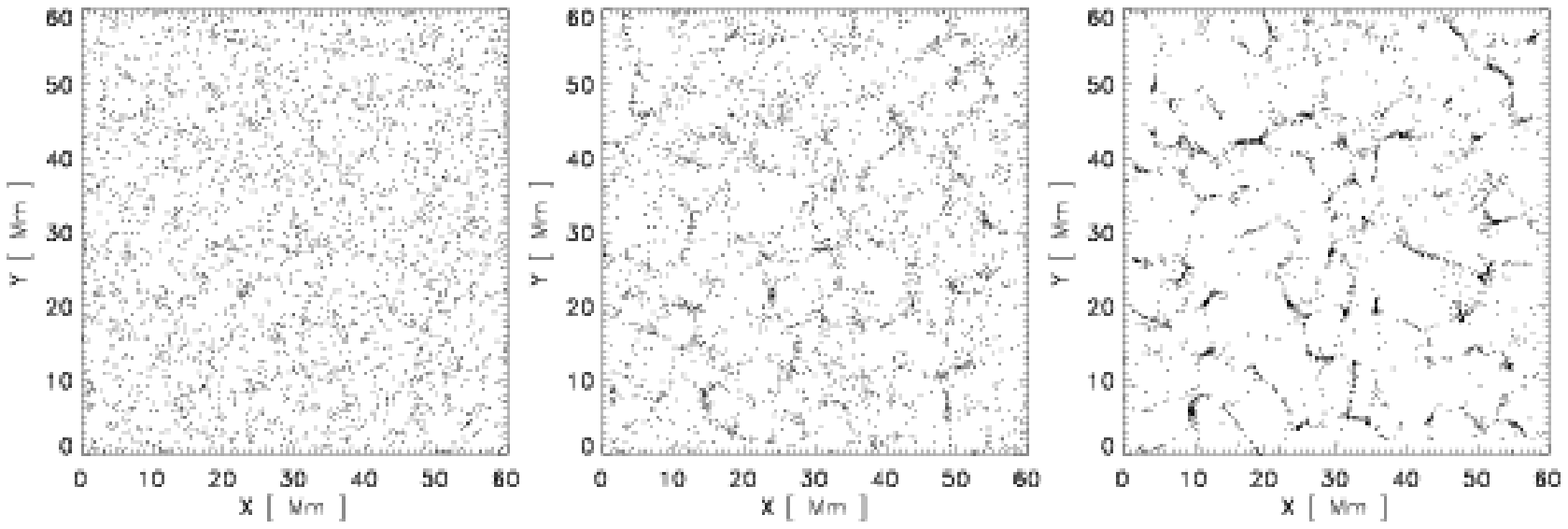}
                \epsscale{1.0}
                \caption{\CAPIII}
                \label{fig:corks}
                \end{figure}
        }

        \newcommand{\FIGIV}{
                \begin{figure}[t]
                \figurenum{4}
                \plotone{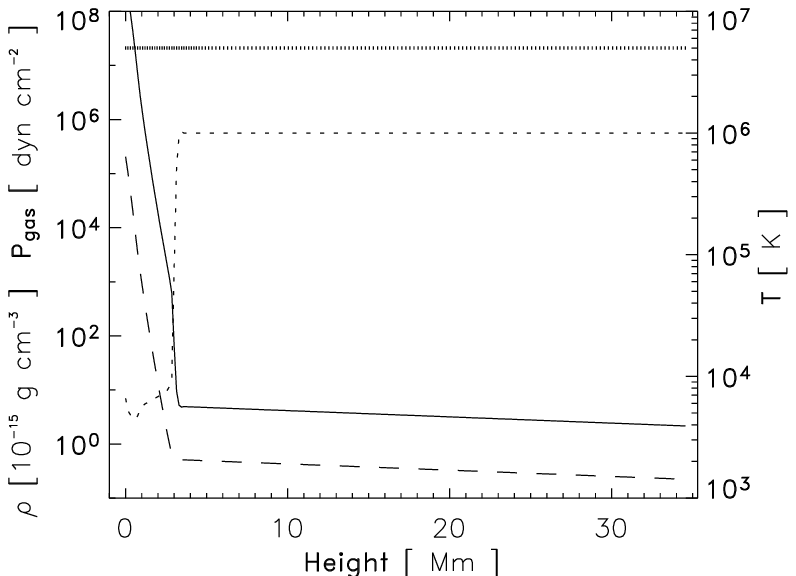}
                \caption{\CAPIV}
                \label{fig:initstrat}
                \end{figure}
        }

        \newcommand{\FIGV}{
                \begin{figure}[t]
                \figurenum{5}
                \plotone{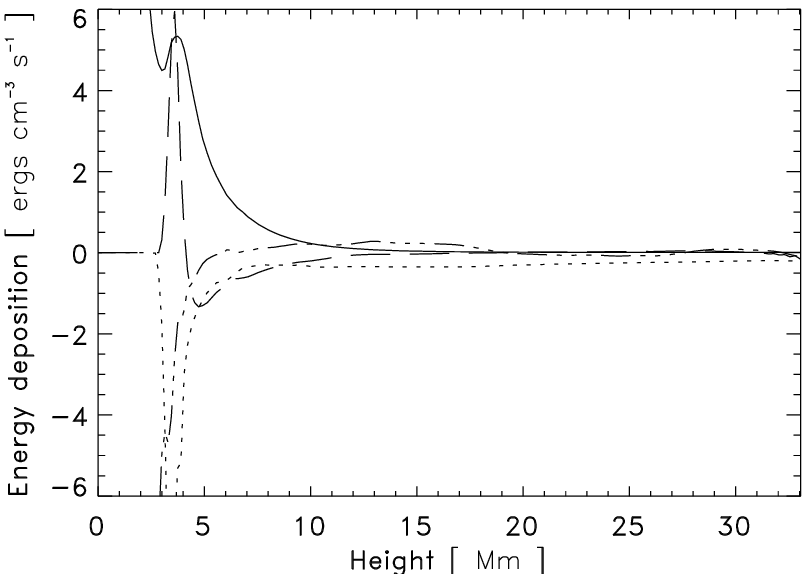}
                \caption{\CAPV}
                \label{fig:balance}
                \end{figure}
        }

        \newcommand{\FIGVII}{
                \begin{figure}[t]
                \figurenum{7}
                \plotone{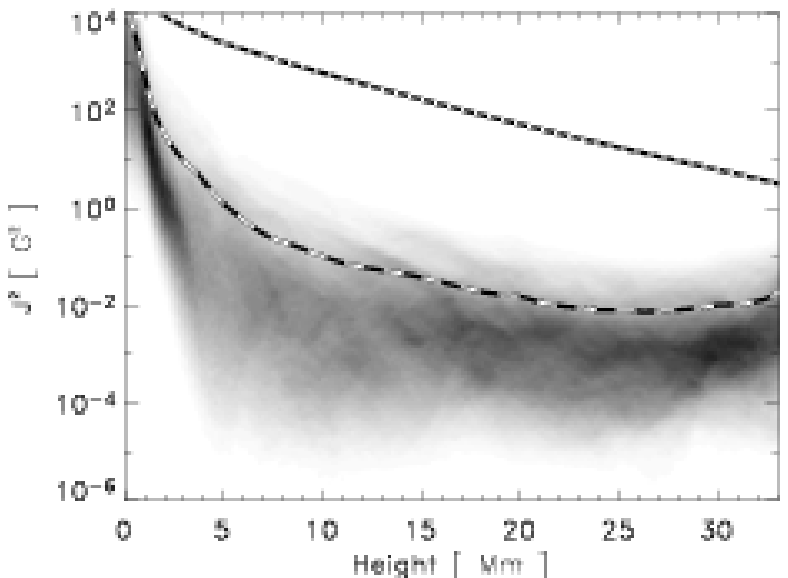}
                \caption{\CAPVII}
                \label{fig:current}
                \end{figure}
        }

        \newcommand{\FIGVI}{
                \begin{figure}[t]
                \figurenum{6}
                \plotone{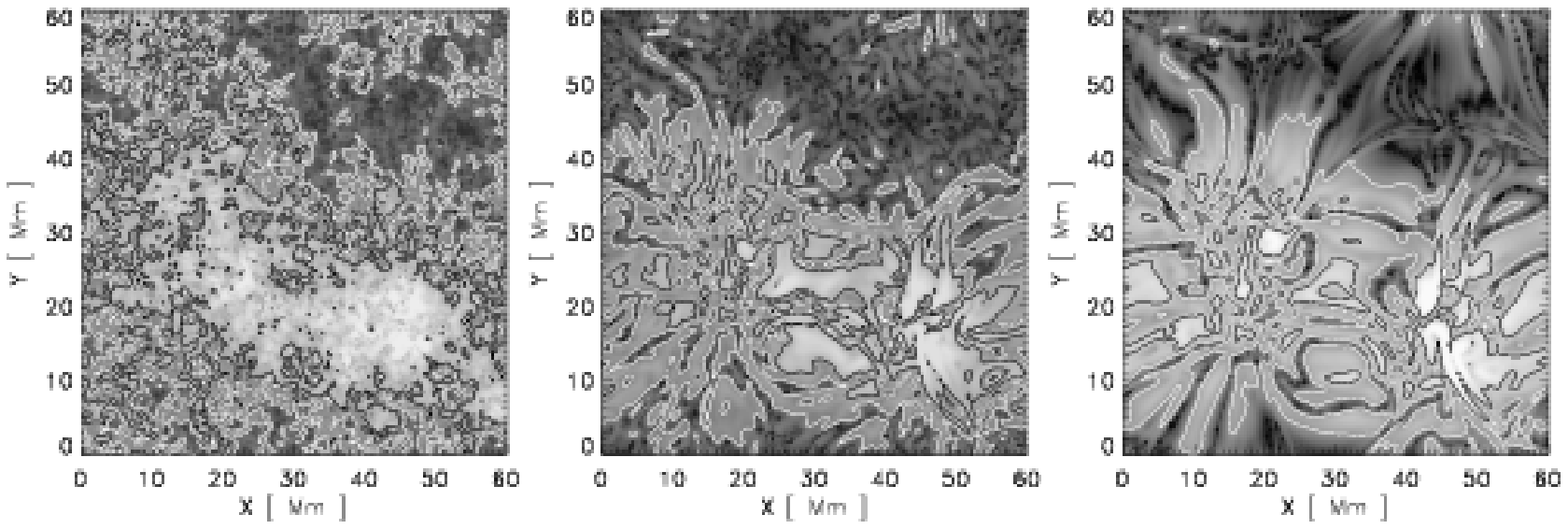}
                \epsscale{1.0}
                \caption{\CAPVI}
                \label{fig:alpha}
                \end{figure}
        }

        \newcommand{\FIGIX}{
                \begin{figure}[t]
                \figurenum{9}
                \plotone{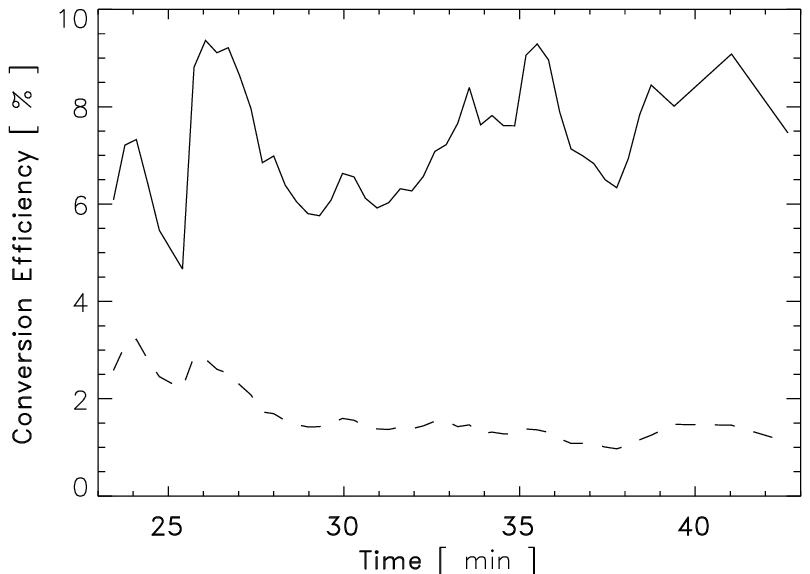}
                \caption{\CAPIX}
                \label{fig:efficiency}
                \end{figure}
        }

        \newcommand{\FIGVIII}{
                \begin{figure}[t]
                \figurenum{8}
                \plotone{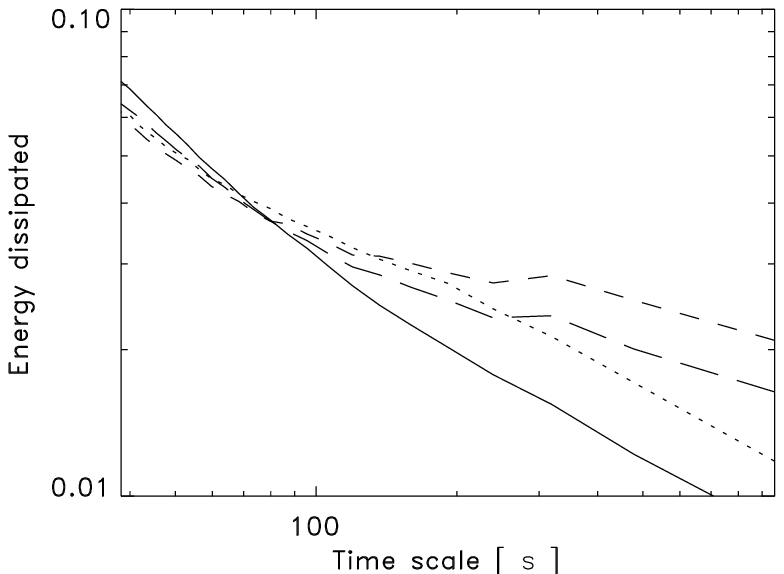}
                \epsscale{1.0}
                \caption{\CAPVIII}
                \label{fig:heat-time}
                \end{figure}
                }

        \newcommand{\FIGX}{
                \begin{figure}[t]
                \figurenum{10}
                \plotone{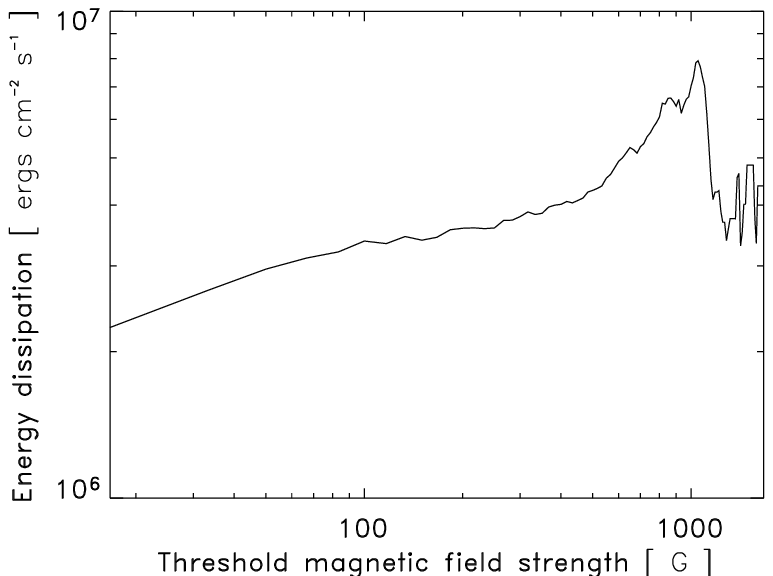}
                \caption{\CAPX}
                \label{fig:totenergy}
                \end{figure}
        }

        \newcommand{\FIGXI}{
                \begin{figure}[t]
                \figurenum{11}
                \plotone{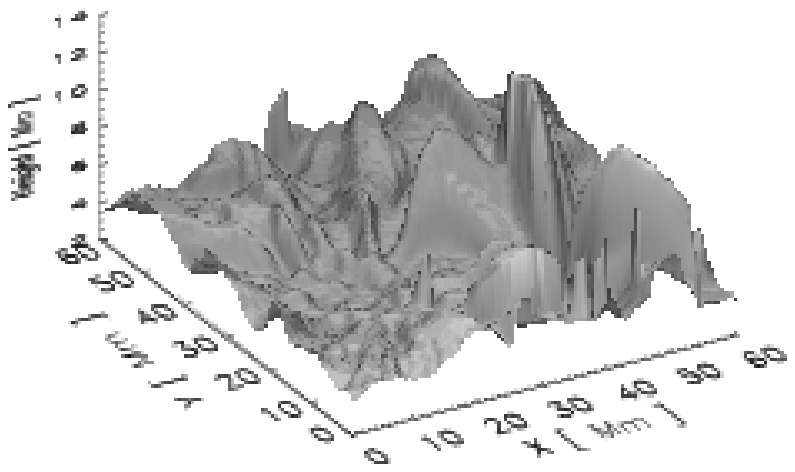}
                \caption{\CAPXI}
                \label{fig:transregion}
                \end{figure}
        }

        \newcommand{\FIGXII}{
                \begin{figure}[t]
                \figurenum{12}
                \plotone{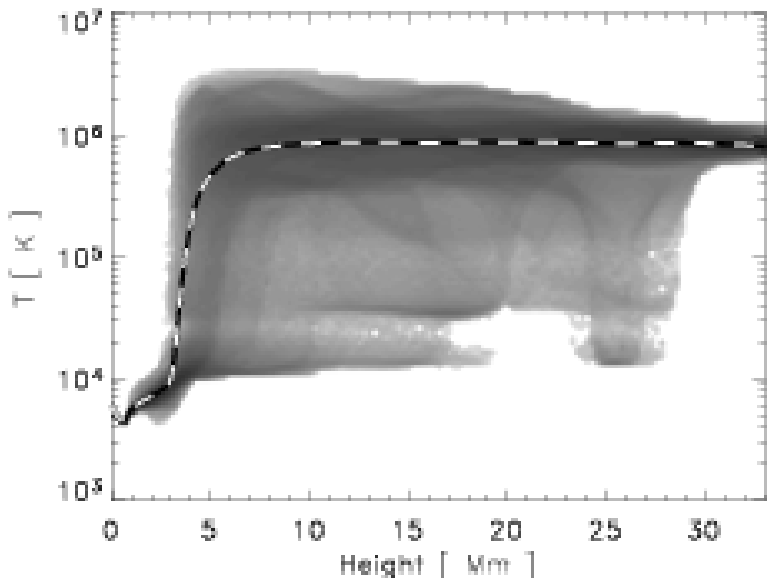}
                \caption{\CAPXII}
                \label{fig:tstrat}
                \end{figure}
        }

        \newcommand{\FIGXIII}{
                \begin{figure}[t]
                \figurenum{13}
                \plotone{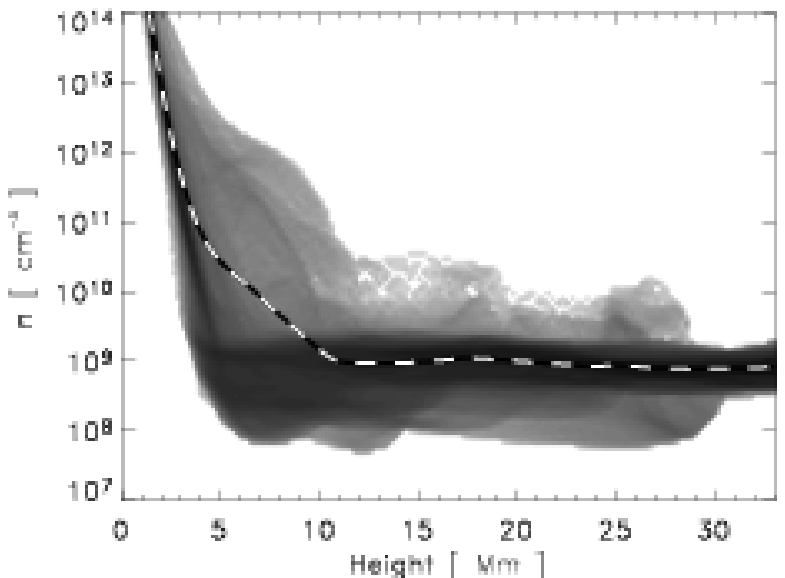}
                \caption{\CAPXIII}
                \label{fig:rhostrat}
                \end{figure}
        }

        \newcommand{\FIGXIV}{
                \begin{figure}
                \figurenum{14}
                \includegraphics[]{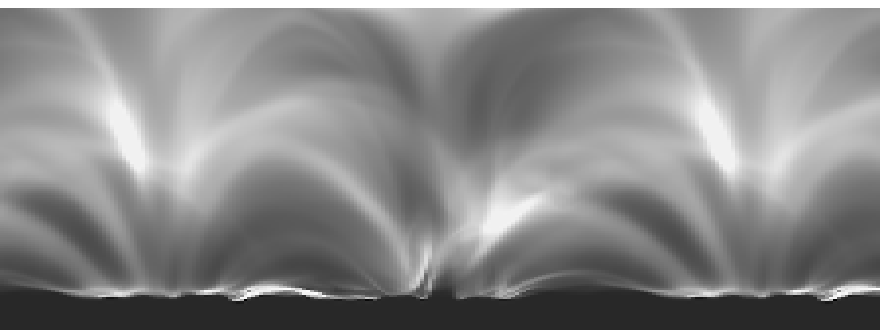}\includegraphics[]{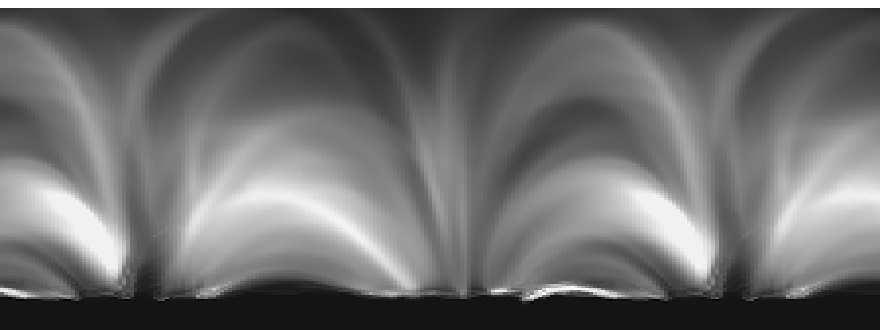}
                \includegraphics[]{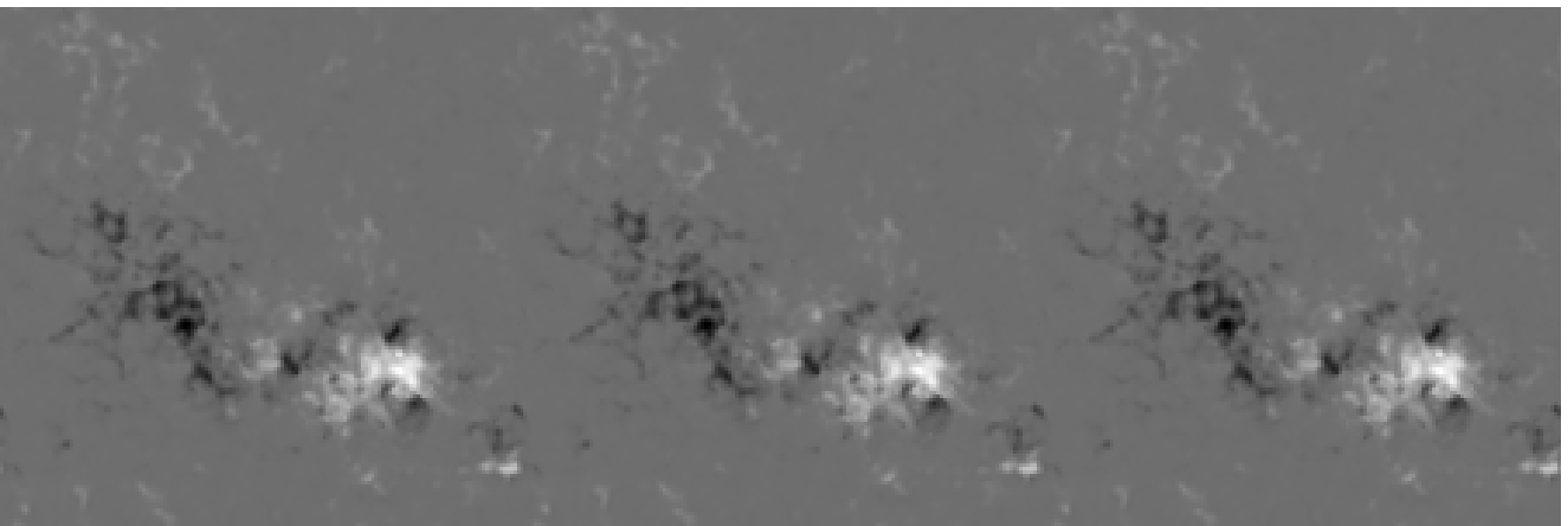}
                \caption{\CAPXIV}
                \label{fig:TRACE}
                \end{figure}
        }
        
\else

        \newcommand{\FIGI}{
                \begin{figure}[t]
                  \figurenum{1}
                  \centerline{\includegraphics[width=8.2cm]{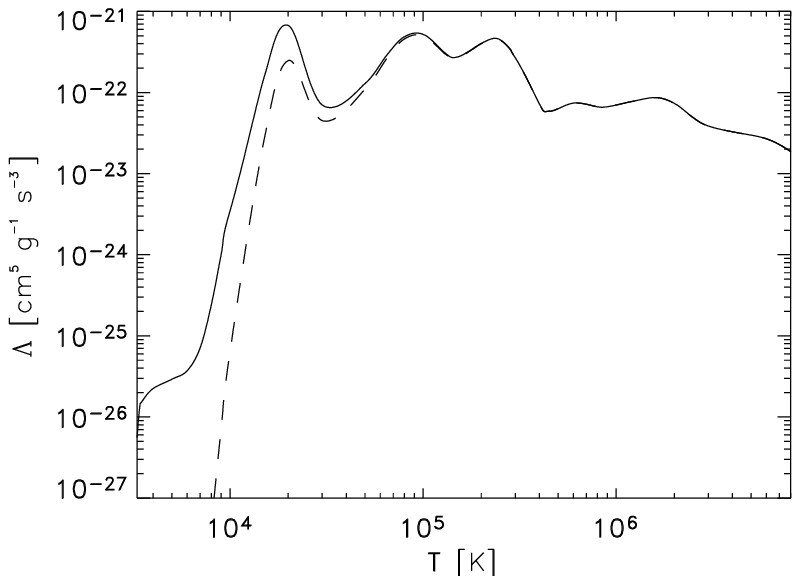}}
                  \caption{\CAPI}
                  \label{fig:lambda}
                \end{figure}
        }

        \newcommand{\FIGII}{
                \begin{figure}[t]
                  \figurenum{2}
                  \centerline{\includegraphics[width=8.2cm]{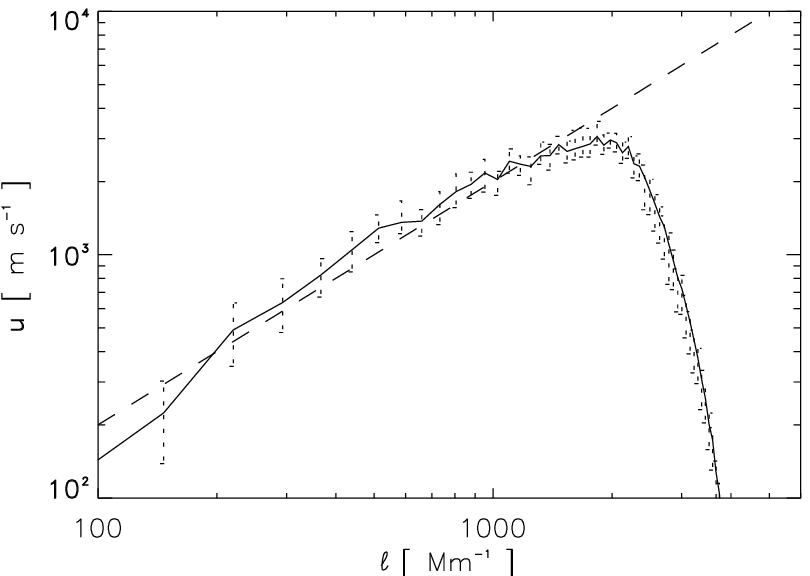}}
                  \caption{\CAPII}  
                  \label{fig:vspectrum}
                \end{figure}
        }

        \newcommand{\FIGIII}{
                \begin{figure*}
                  \figurenum{3}
                  \centerline{\includegraphics[]{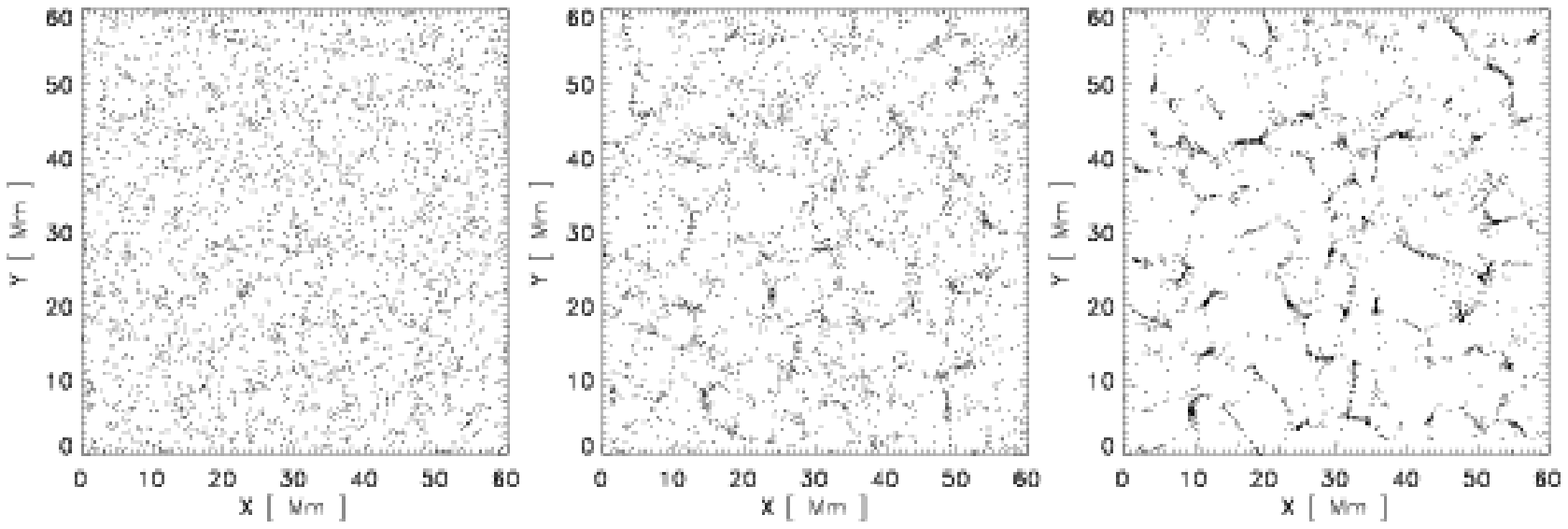}}
                  \caption{\CAPIII}
                  \label{fig:corks}
                \end{figure*}   
        }

        \newcommand{\FIGIV}{
                \begin{figure}[t]
                  \figurenum{4}
                  \centerline{\includegraphics[width=8.2cm]{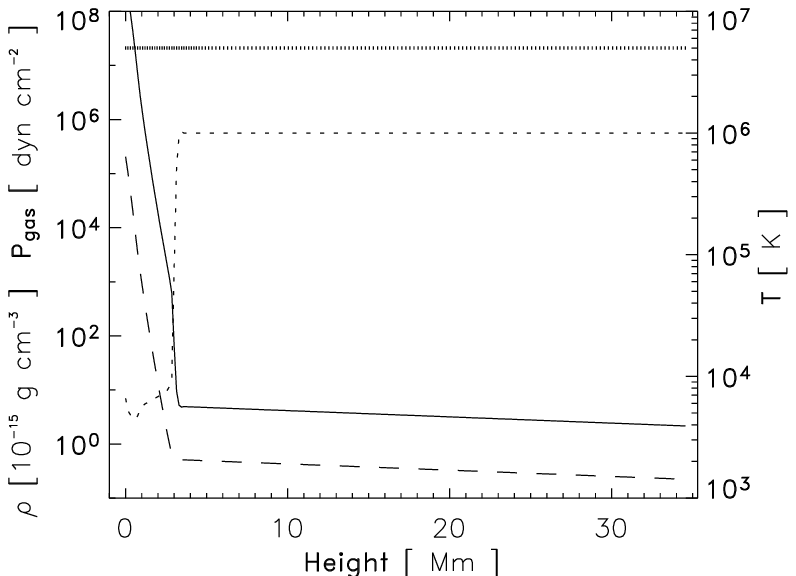}}
                  \caption{\CAPIV}
                  \label{fig:initstrat}
                \end{figure}
        }

        \newcommand{\FIGV}{
                \begin{figure}[t]
                  \figurenum{5}
                  \centerline{\includegraphics[width=8.2cm]{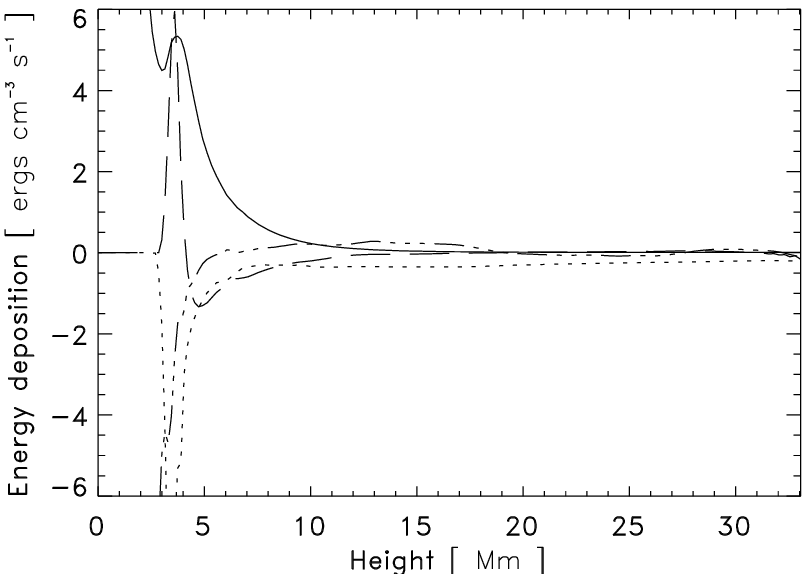}}
                  \caption{\CAPV}
                  \label{fig:balance}
                \end{figure}
        }

        \newcommand{\FIGVII}{
                \begin{figure}[t]
                  \figurenum{7}
                  \centerline{\includegraphics[width=8.2cm]{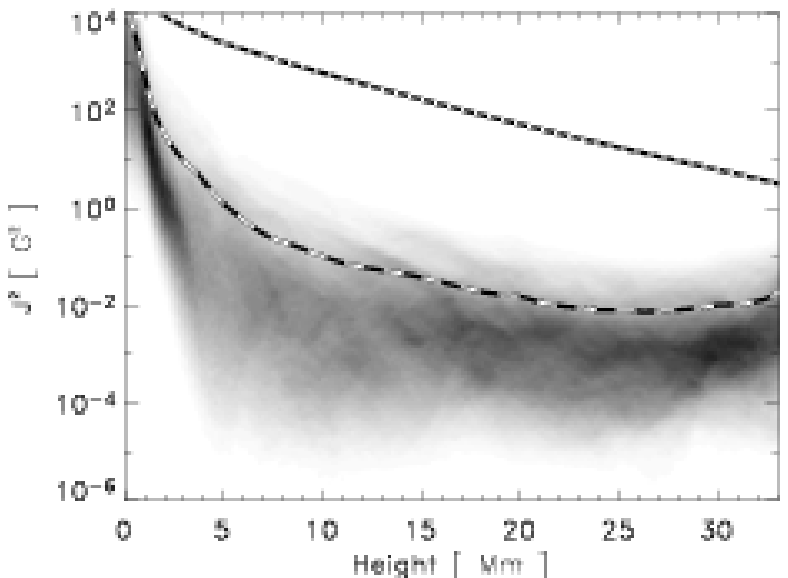}}
                  \caption{\CAPVII}
                  \label{fig:current}
                \end{figure}
        }

        \newcommand{\FIGVI}{
                \begin{figure*}
                  \figurenum{6}
                  \centerline{\includegraphics[]{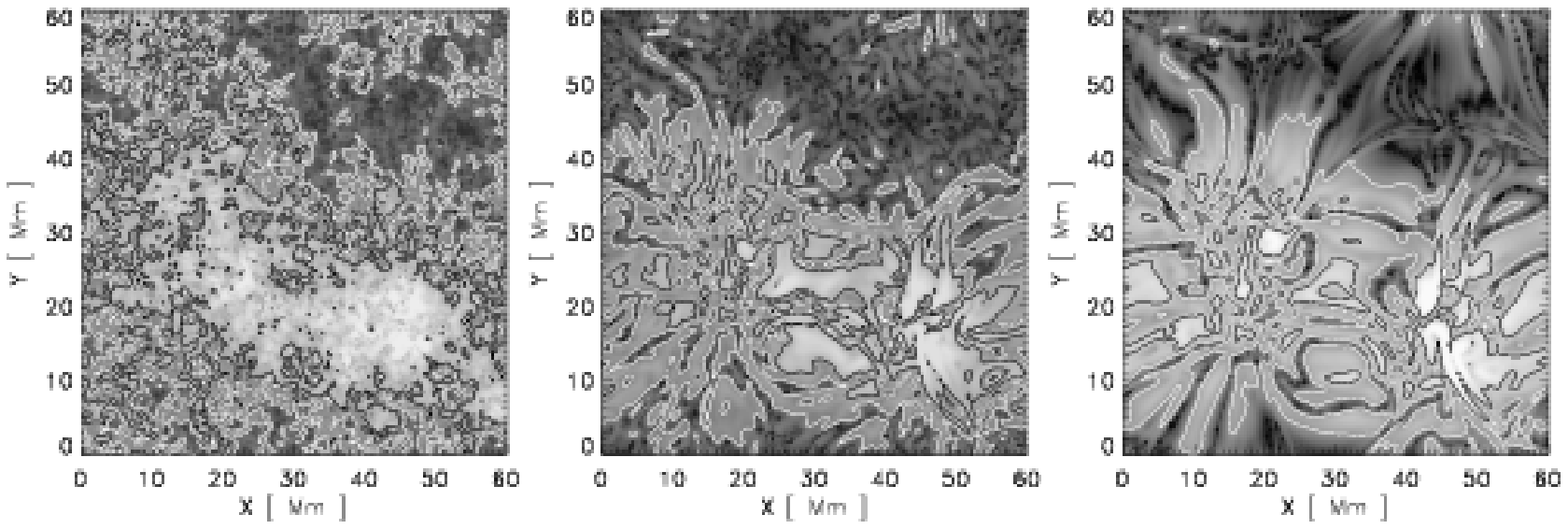}}
                  \caption{\CAPVI}
                  \label{fig:alpha}
                \end{figure*}
        }

        \newcommand{\FIGIX}{
                \begin{figure}[t]
                  \figurenum{9}
                  \centerline{\includegraphics[width=8.2cm]{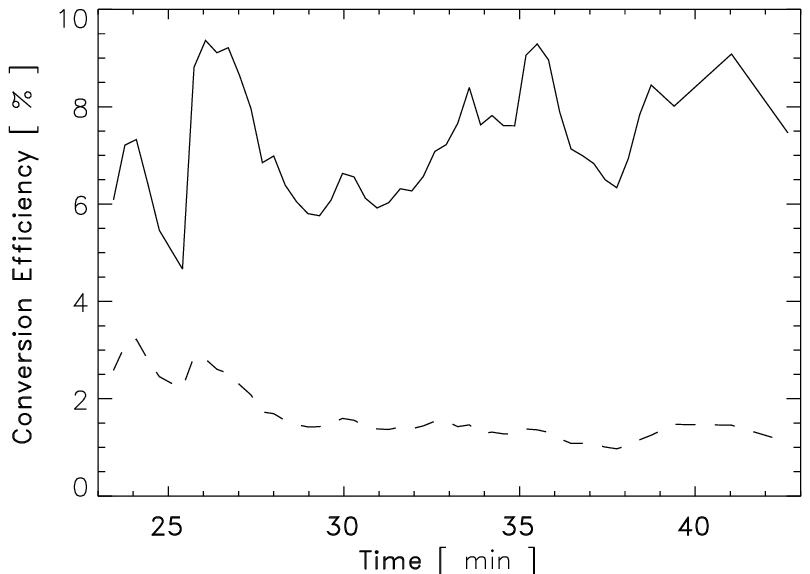}}
                  \caption{\CAPIX}
                  \label{fig:efficiency}
                \end{figure}
        }

        \newcommand{\FIGVIII}{
                \begin{figure}[t]
                  \figurenum{8}
                  \centerline{\includegraphics[width=8.2cm]{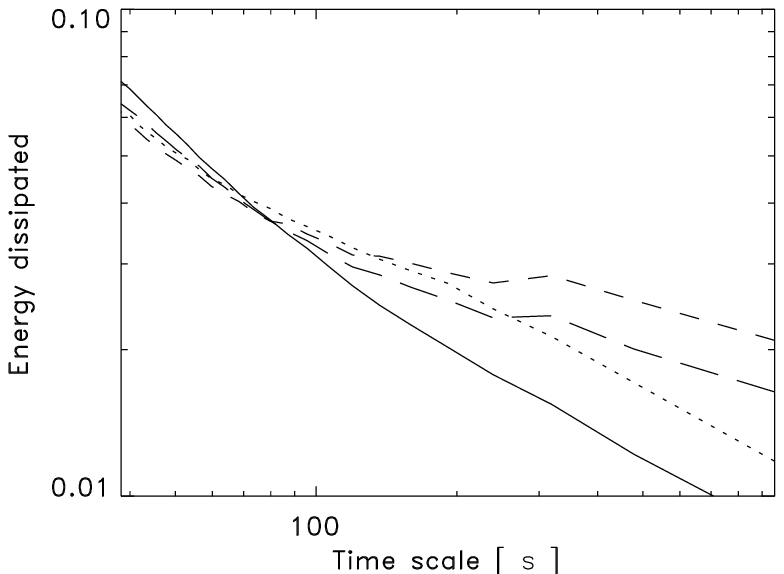}}
                  \caption{\CAPVIII}
                  \label{fig:heat-time}
                \end{figure}
        }

        \newcommand{\FIGX}{
                \begin{figure}[t]
                  \figurenum{10}
                  \centerline{\includegraphics[width=8.2cm]{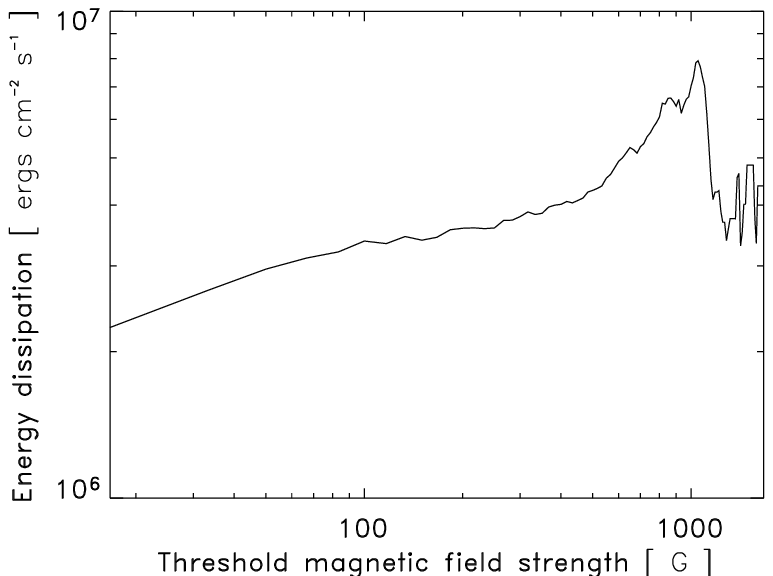}}
                  \caption{\CAPX}
                  \label{fig:totenergy}
                \end{figure}
        }

        \newcommand{\FIGXI}{
                \begin{figure}[t]
                  \figurenum{11}
                  \centerline{\includegraphics[width=8.2cm]{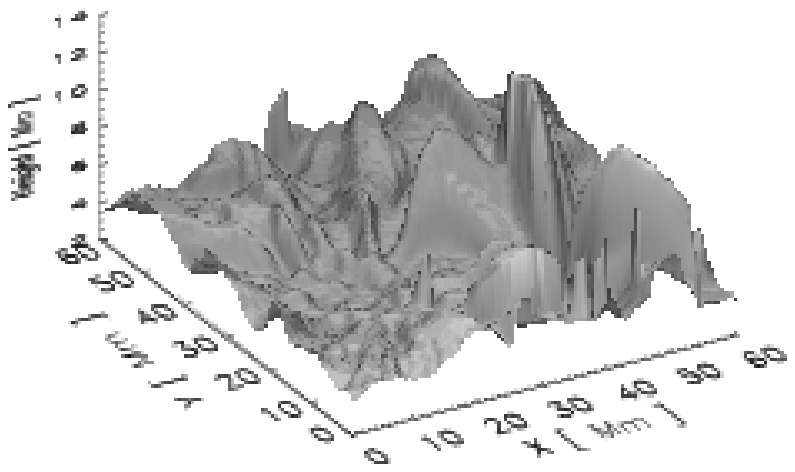}}
                  \caption{\CAPXI}
                  \label{fig:transregion}
                \end{figure}
        }

        \newcommand{\FIGXII}{
                \begin{figure}[t]
                  \figurenum{12}
                  \centerline{\includegraphics[width=8.2cm]{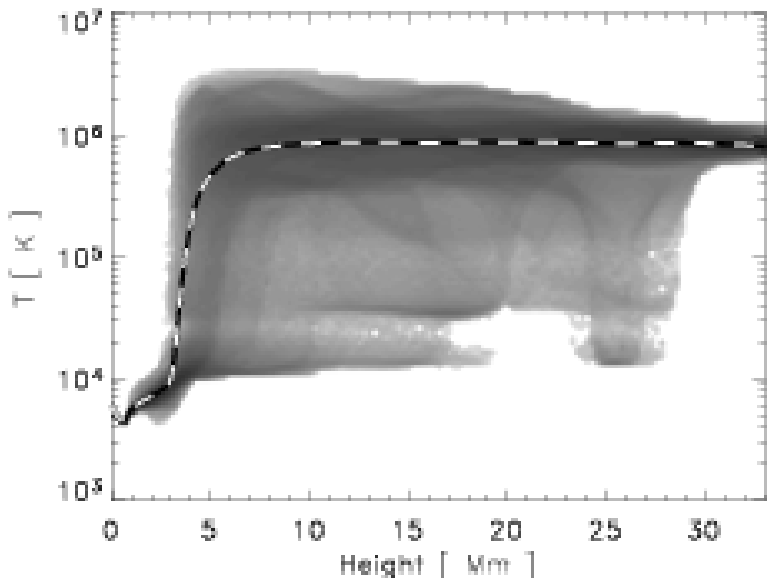}}
                  \caption{\CAPXII}
                  \label{fig:tstrat}
                \end{figure}
        }

        \newcommand{\FIGXIII}{
                \begin{figure}[t]
                  \figurenum{13}
                  \centerline{\includegraphics[width=8.2cm]{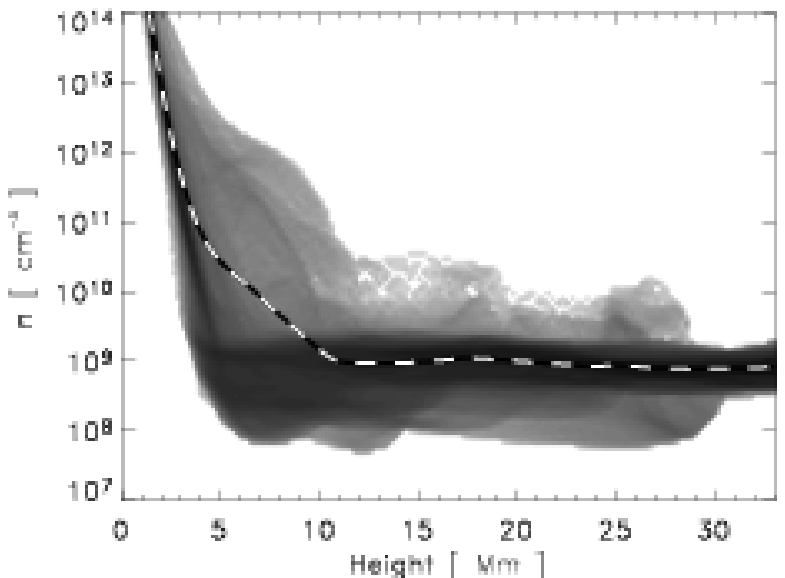}}
                  \caption{\CAPXIII}
                  \label{fig:rhostrat}
                \end{figure}
        }

        \newcommand{\FIGXIV}{
                \begin{figure*}
                  \figurenum{14}
                  \hspace*{0.25cm}\includegraphics[]{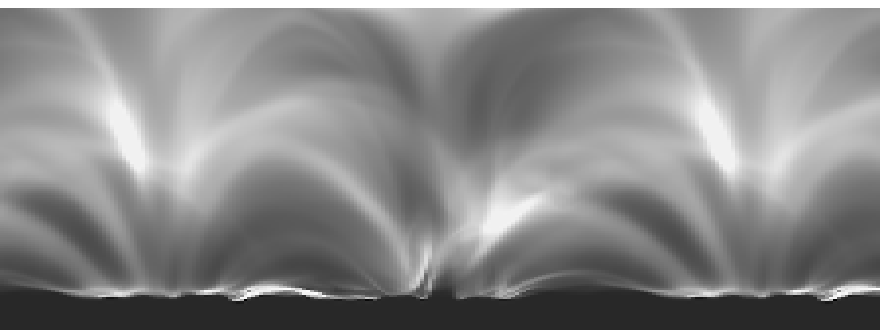}\includegraphics[]{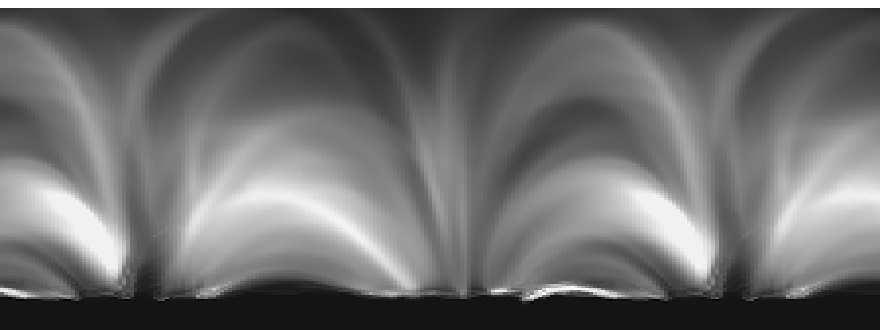}
                  \hspace*{0.25cm}\includegraphics[]{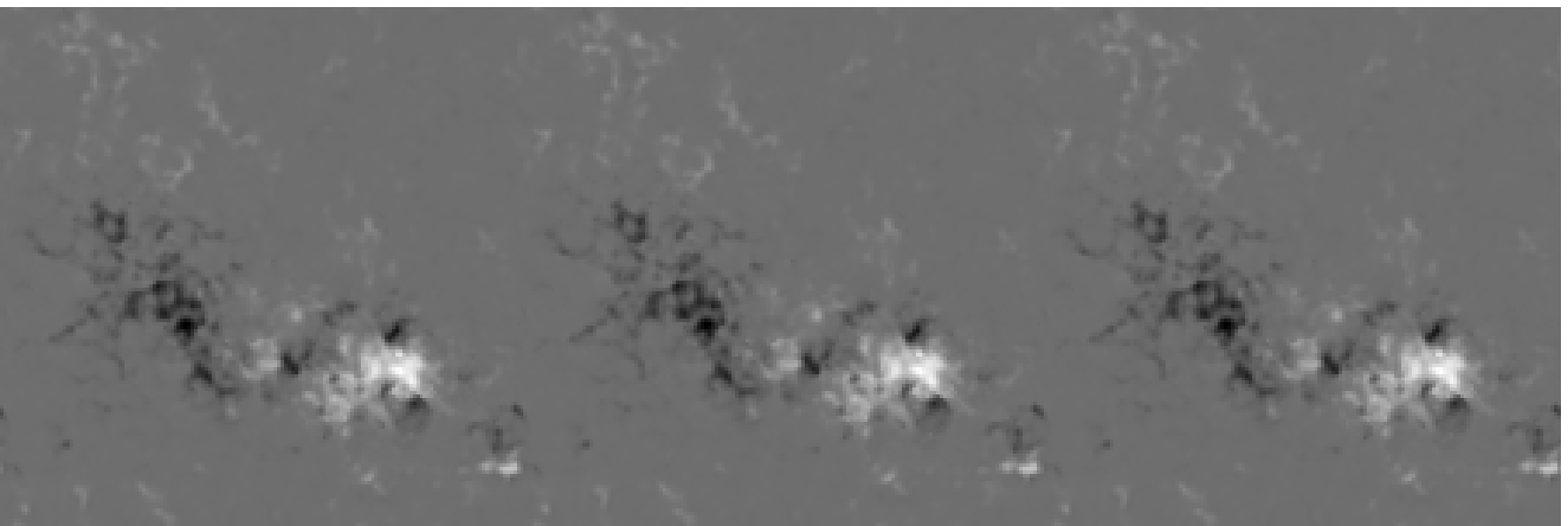}\hspace*{0.5cm}
                  \caption{\CAPXIV} 
                  \label{fig:TRACE}
                \end{figure*}
        }

\fi

\newcommand{\un}[1]{\ensuremath{\,{\rm{#1}}}}
\newcommand{\bea}{\begin{eqnarray*}}
\newcommand{\eea}{\end{eqnarray*}}
\newcommand{\bean}{\begin{eqnarray}}
\newcommand{\eean}{\end{eqnarray}}
\newcommand{\Div}{\vec{\nabla}\cdot}
\newcommand{\Grad}{\vec{\nabla}}
\newcommand{\Rot}{\vec{\nabla}\times}
\newcommand{\ddt}[1]{\frac{\partial #1}{\partial t}}
\newcommand{\BB}{\vec{B}}
\newcommand{\JJ}{\vec{J}}

\newcommand{\Fig}[1]{Fig.\ \ref{#1}}
\newcommand{\Figure}[1]{Figure \ref{#1}}

\begin{document}

\title{An Ab Initio Approach to the Solar Coronal Heating Problem}

\ifnum\astroph=1
        \author{B.V. Gudiksen and {\AA}. Nordlund}
\else
        \author{Boris Vilhelm Gudiksen\altaffilmark{1}}
        \affil{The Institute for Solar Physics of the Royal Swedish
        Academy of Sciences}
        \affil{Albanova University Center, Stockholm Observatory\\
        10691 Stockholm\\ Sweden} 
        \email{boris@astro.su.se}
        \and
        \author{{\AA}ke Nordlund}
        \affil{Astronomical Observatory}
        \affil{NBIAfG, Copenhagen University}
        \affil{{\O}ster Voldgade 3\\1350 Copenhagen K\\Denmark}
        \email{aake@astro.ku.dk}
        \altaffiltext{1}{Now at Inst. for Theoretical Astrophysics,
        University of Oslo, email: boris@astro.uio.no} 
\fi

\begin{abstract}
We present an {\em{ab initio}} approach to the solar coronal heating
problem by modelling a small part of the solar corona in a 
computational box using a 3D MHD code including realistic physics. The
observed solar granular velocity pattern  
and its amplitude and vorticity power spectra, as reproduced by a weighted 
Voronoi tessellation method, are used as a boundary condition that 
generates a Poynting flux in the presence of a magnetic field.
The initial magnetic field is a potential extrapolation of a SOHO/MDI high
resolution magnetogram, and a standard stratified atmosphere is used
as a thermal initial condition. 
Except for the chromospheric temperature structure,
which is kept fixed, the initial conditions are quickly forgotten
because the included Spitzer conductivity and radiative cooling
function have typical timescales much shorter than the time
span of the simulation. After a short initial start
up period, the magnetic field is able to dissipate $3-4\times 10^6
\un{ergs\,cm^{-2}\,s^{-1}}$ in a highly intermittent corona,
maintaining an average temperature of $\sim 10^6$ K, at coronal 
density values for which emulated images
of the Transition Region And Coronal Explorer(TRACE) 171 and 195 {\AA}
pass bands reproduce observed photon count rates.
\end{abstract}
\keywords{Sun: corona -- Sun: magnetic fields -- MHD}

\section{Introduction}
The heating mechanism at work in the solar corona has puzzled
researchers for more than six 
decades. Several heating mechanisms have 
been proposed. One of the earliest models involved accretion of
interstellar matter coupled with convection in the 
chromosphere \citep{Bondi+etal47}, but with improved measurements two
main groups of heating models emerged: Wave (AC) heating and electric
current (DC) heating. These
mechanisms move energy from the photospheric kinetic energy reservoir
to the internal heat reservoir in the corona through the magnetic
field. AC heating \citep[already proposed by][]{Alfven47} depends 
on the magnetic field being moved around  
in the solar photosphere faster than the disturbances can propagate
through the whole magnetic loop, i.e.\ faster than the Alfv{\'e}n
crossing time. Only torsional Alfv{\'e}n waves can reach the corona,
while magneto-sonic wave modes are diffracted and dissipated due to
the strong wave speed gradient in the chromosphere/transition
region. Waves have been detected both in the near Sun corona
\citep[see for instance][]{Aschwanden87,Ofman+Davila97} as well 
as in the solar wind by a number of spacecrafts such as the Helios
satellites \citep{Neubauer+Musmann77} so there is firm evidence for
their existence, but not of their ability to supply enough
heat. Alfv{\'e}n waves do not easily dissipate in the corona in the
absence of phase mixing \citep{Heyvaerts+Priest83} or resonant
absorption \citep{Davila87}, which require strong phase speed gradients to
be efficient.

DC heating relies on motions in the
photosphere that change on time scales longer than Alfv{\'e}n crossing times, 
making the magnetic field remain close to an equilibrium state. 
The DC heating mechanism dissipates energy
through conventional Joule heating \citep[proposed by][]{Parker72},
or reconnection heating \citep[proposed by][based on X--ray
observations]{Glencross+etal74} dissipated through flares of all
sizes, including nano--flares, or through a hierarchy of current
sheets that may encompass both of the above processes 
\citep{Galsgaard+Nordlund96}.
Until recently, nano--flares appeared
to be the most promising candidate, but it now seems that the
energy dissipated in the observed flare distribution, extrapolated
to a cut-off at nano--flare energies, is too small
\citep[][]{Aschwanden+etal00a,Parnell+Jubb00,Aschwanden+Parnell02,
Aschwanden+Charb.02}. DC heating has received considerable attention  
since it was proposed some thirty years ago \citep[e.g.][ to mention
just a few]{Parker72,Parker83,Sturrock+Uchida81, vanBallegooijen86,
Mikic+etal89,Heyvaerts+Priest92, Longcope+Sudan94,
Galsgaard+Nordlund96,Hendrix+etal96, Gomez+etal00}. It has been
established that the DC heating mechanism is feasible, but not to what
extent it can provide enough energy to heat the solar corona. 

The main problem
in modelling solar dissipative processes is the length scales
involved. The dissipative length scale is of the order of $1\un{m}$
for e.g.\ a flare with a time scale of the order $100
\un{s}$, while the large scale magnetic field distribution has a scale
of many $\un{Mm}$, making fully resolved 3D simulations of such
processes practically impossible. However, the theoretical results of
\cite{Parker79} and the numerical results of \cite{Galsgaard+Nordlund96} and
\cite{Hendrix+etal96} have shown that it is not necessary to resolve the
dissipative length scales because the
total dissipation only depends very weakly on the resolution,
and if anything increases with
increasing resolution.  Results from low resolution modelling of 
dissipative processes may thus be used to obtain estimates, or at least
establish {\it{lower}} limits on the dissipated energy. This 
provides the basis for trying to create numerical {\em ab initio}
simulations of the solar corona, with essentially
no free parameters.  As discussed in more detail below, the initial 
and boundary conditions that we use are {\em minimal assumptions}, 
in the sense that including more detailed
conditions would produce more heating. This allows us to give
a firm answer to whether DC heating can provide the energy required
and thus which heating mechanism is the principal one
in the solar corona.

Initial results of this effort were presented in
\citet{Gudiksen+Nordlund02} and here we present further
developments including a realistic photospheric velocity driver and
a non uniform mesh, making it possible to resolve the density drop in the
transition region and expand the simulated volume. 

\section{Model}
We solve the fully compressible MHD equations on a non--uniform staggered
mesh, in the form
\bean
\ddt{\rho}&=&-\Div \rho \vec{u}\\
\ddt{\rho \vec{u}}&=&-\Div \left(\rho \vec{u}
\vec{u}-\tau\right)-\Grad 
P+\vec{J}\times\vec{B}+\rho \vec{g}\\
\mu \vec{J}&=&\Rot{\vec{B}}\\
\vec{E}&=&\eta\vec{J}-\vec{u}\times\vec{B}\\
\ddt{\vec{B}}&=&-\Rot{\vec{E}}\label{eq:induction}\\
\ddt{e}&=&-\Div e\vec{u}-P\Div \vec{u} \nonumber\\
&&-\vec{\nabla}\cdot
\vec{F}_{\mathrm{Spitzer}}+Q_{\mathrm{visc}}+Q_{\mathrm{Joule}}-n_{ion}n_{e}\Lambda ~,
\eean
where $\rho$ is the mass density, $\vec{u}$ the velocity vector,
$\tau$ the viscous stress tensor, $P$ the gas pressure,
$\vec{J}$ the electric
current density, $\vec{B}$ the magnetic flux density, $\vec{g}$ the
gravitational acceleration, $\mu$ the vacuum permeability, $\vec{E}$
the electric field strength, $\eta$ the magnetic diffusivity, $e$ the internal
energy per unit volume, $\vec{F}_{\mathrm{Spitzer}}$ is the energy flux due to
the Spitzer conductivity \citep{Spitzer56} along the magnetic field,
$Q_{\mathrm{visc}}$ is the 
viscous heating, $Q_{\mathrm{Joule}}$ is the Joule heating, and
$\Lambda$ is the cooling function for the optically thin coronal
plasma with $n_{ion}$ and $n_e$ being the number density of ions and
electrons.

\ifnum\astroph=1
\FIGI
\fi

The code is explicit and, because of the staggered mesh method, 
conserves the magnetic divergence to machine precision. The staggered
mesh method means that the physical variables are not 
aligned in space. This is an advantage for the derivative operators,
but has the downside that when physical variables at different
locations are involved interpolation is needed to realign the
variables. The code uses 6'th order derivative operators and 5'th order
interpolation operators, together with a 3'rd order Runge-Kutta
time stepping method with variable time step. The code has been
extensively subjected to several standard tests as well as used in
several physical regimes, for instance 3-D 
turbulence and magneto-convection \citep{Nordlund+etal94}, magnetic dissipation
\citep{Galsgaard+Nordlund96}, buoyant magnetic flux
tubes \citep{Dorch+Nordlund98}, emergence of flux ropes through the
solar convection zone \citep{Dorch+etal99}, and supersonic MHD-turbulence 
\citep{Padoan+Nordlund99}.

The computational box of $60\times60\times37\un{Mm^3}$ is
resolved on a grid of $150^3$ grid points where the vertical scale is
non--uniform. The vertical axis starts at the photosphere, with high 
resolution ($\sim 0.15 \un{Mm}$) in
order to resolve the small pressure scale height in the photosphere and
chromosphere. Above the transition region
the vertical resolution gradually approaches 0.25 Mm. 

\subsection{Radiative cooling}
The cooling function includes the most important transitions for H,
He, C, O, Ne, Fe and bremsstrahlung, with a cut off at low
temperatures to avoid catastrophic cooling in the chromosphere (see
\Fig{fig:lambda}). The cooling function is based on the ionisation and
recombination rates given by
\citet{Arnaud+Rothenflug85,Shull+vanSteenberg82} and using the
collisional excitation rates found through the HAO-DIAPER 
Atom data package \citep{Judge+Meisner94}. The ions are treated by
assuming ionisation equilibrium making it possible to derive radiative
losses as a function of electron temperature.

Further down in the atmosphere, where the radiation is optically thick
for at least some wavelengths, the cooling processes cannot be treated
with a simple cooling function. The cooling function shown in
\Fig{fig:lambda}(full) uses a cutoff proportional to
$\exp{\left(-\tau\right)}$ with
the optical depth $\tau \propto P_{gas}$. In this simulation, the lower
part of the chromosphere and the photosphere are instead kept near a
fixed average temperature profile using a Newtonian cooling mechanism
that forces the local temperature towards a preset function of height
on a timescale of about 0.1 s in the photosphere. The time scale
increases with decreasing density as $\rho^{-1.67}$, and the effect
thus becomes negligible in the corona.

\ifnum\astroph=1
\FIGII
\fi

\ifnum\astroph=1
\FIGIII
\fi

\subsection{Boundary conditions}\label{sec:boundaries}
The vertical boundary conditions are handled by using ghost zones, while the
box is periodic in the horizontal directions. The density is
extrapolated into the ghost zones, and to prevent heat from leaking
into or out of the box, the temperature
gradient is set to zero at the upper boundary. The upper boundary is
relaxed towards the average temperature at the upper boundary with a
typical time scale of about 30 s. The vertical velocity is zero on the
boundaries, while the horizontal velocity on the lower boundary is set
by a driver (See \S \ref{sec:driver}). At the upper boundary the
vertical gradient of the horizontal velocity is set to zero and we use
a potential field extrapolation for the magnetic field in the ghost zones.

\subsection{The Voronoi tessellation driver}\label{sec:driver}
The solar photospheric velocity field is characterised by granular
structures with a range of sizes, from granules with a typical size
of $\sim 1$ Mm to super-granules of size $\sim 20-30$ Mm and even
bigger giant cells \citep{Simon+Weiss68,Hathaway+etal00}. These
structures have typical velocities that scale
with wavenumber roughly as $u \propto k$ and have
turnover times of the order $\tau \propto k^{-2}$, all the
way from beyond super-granular scales to the velocity maximum at
granular scales
\citep{Stein+Nordlund98,Hathaway+etal00,Shine+etal00}. The velocity
field injects energy into the magnetic field through a Poynting flux,
making a realistic reproduction important.

Our goal was
to create a driver that would match these statistical features of the
solar velocity spectrum, as well as the geometrical features of the
granular pattern.
The velocity driver is partially based on the theory of {\em{Voronoi
tessellation}} \citep[see for instance][]{voronoibook} specifically
multiplicatively weighted Voronoi tessellation, inspired by
\citet{Schrijver+etal97} who showed that this method gives a
good fit to observations of the granulation pattern. The
theory applied to this case provides a way to split a 2-D surface into
tiles, each representing a granule. To do this one has to chose a
number of generator points, each with a weight $w_i$ (where $i$
represents a specific tile) positioned at $\vec{x}_i$. Tile $i$ then
occupies an area where the following inequality is satisfied,
\bean
\frac{w_i}{ \mid \vec{x} - \vec{x}_i \mid} \ge \frac{w_j}{ \mid
\vec{x}-\vec{x}_j \mid }
\hspace{1cm}\textrm{for all }\, j\neq i ~.
\label{eq:vor-tes}
\eean

To include the
intergranular lanes, we increased the complexity of
Eq.\ref{eq:vor-tes},
\bean
\frac{w_i(1-l_{ig})}{\mid\vec{x}-\vec{x}_i\mid}\ge\frac{w_j}{\mid\vec{x}-\vec{x}_j\mid} 
<
\frac{w_i(1+l_{ig})}{\mid\vec{x}-\vec{x}_i\mid}  ~.
\eean

When both conditions are true $\vec{x}$ is in an intergranular lane,
and where only the left part of the 
inequality is true $\vec{x}$ is in granule $i$. $l_{ig}$ is a number
between zero and one and can be
adjusted to make the intergranular area roughly 33-45\% of the 
total area \citep[][]{Stein+Nordlund98} based on
velocity maps. The relative area of the intergranular lanes can be  
estimated by assuming perfectly round granules,
\bean
\frac{A_{ig}}{A_{tot}}=2l_{ig}\frac{(1-l_{ig})}{(1+l_{ig})} ~.
\eean

\subsubsection{The initial tessellation}
If the generator points were placed randomly there would be 
both unusually large and unusually small granules generated, 
because the density  of generator points would be low in 
some places and high in others. Since observed
granules are rather uniform in size, only a certain
range of sizes of fully grown granules can be accepted. This places
restrictions on the placement of the generator points. 
The tessellation is built up by keeping a
list of the weights at each position, independent of whether it is
in an intergranular lane or not, updating the list as more
granules are placed. In this way a weight function with a value at
each position is generated,
\bean
\mathcal{W}(\vec{x})&=&\frac{w_i}{\mid\vec{x}-\vec{x_i}\mid}\hspace{0.5cm}\textrm{where}\nonumber\\
&&\frac{w_i}{\mid\vec{x}-\vec{x}_i\mid}\ge\frac{w_j}{\mid\vec{x}-\vec{x}_j\mid}\hspace{0.5cm}\textrm{for
  all} \,j\,\neq i ~.
\eean

To generate a new granule, its generator point is placed where the
weight function has a value lower than a certain threshold
$\mathcal{W}_{\mathrm{thresh}}$. This allows more generator points to
be placed only until the weight 
function is above the threshold at all positions. By choosing an appropriate
threshold carefully the density of generator points is such that
granules of the right size are created. 

The velocity vector points radially away from the generator points
inside the granules, 
and along the granular boundaries in the intergranular lanes.
The horizontal velocity profile inside meso-granules was modelled by
\citet{Simon+Weiss89} by fitting a simple model to observations done
with the SOUP filter on Spacelab 2. The horizontal velocity profiles
were successfully fitted with a radial velocity profile of the form
$\vec{v}_h(r)=r\exp{(-(r/r_0)^2)}$. We found a better fit to the solar velocity
power spectrum by using $\vec{v}_h(r)=r^2\exp{(-(r/r_0)^2)}$,
giving a flatter profile in the centre of the granules. In the
intergranular lanes the velocity is a sum of the velocity of the two
granules adjacent to the intergranular lane. Initially this creates
stagnation points in the intergranular lanes, on lines connecting the
generator points, but such stagnation points disappear when multiple 
velocity fields are added (see \S\ref{sec:time_evo}).

\subsubsection{Time evolution}\label{sec:time_evo}
The tessellation is made time dependent by making the weights of the
generator points time dependent. We have chosen to give 
the weights of the generator points a time dependence specified by
\bean
w_i(t)=W_i
\exp{\left(-\left(\frac{t-t_{0,i}}{\tau_i}\right)^q\right)} ~,
\eean
where $W_i$ is a constant giving the maximum weight, $t_{0,i}$ is the
time of maximum weight, and 
$q$ is an even integer, indirectly controlling the growth time of the
granule. We bring the
granule to life when $w_i/dx > \mathcal{W}_{\mathrm{thresh}}$ where $dx$ is the grid resolution, which effectively sets $t_{0,i}$ when $W_i$ is
chosen. When $t > t_{0,i}$ the granule is decaying and it is removed when
$w_i/d\vec{x} < \mathcal{W}_{\mathrm{thresh}}$. 

When the weight function
$\mathcal{W}(\vec{x})$ falls below the threshold in a large connected
region, each new granule covers only a small area,
causing many small granules to appear, thus making the area 
crowded with small granules of approximately equal size. This problem is solved
by creating a binary function $\mathcal{A}(\vec{x},t)$, which is set to
one within 80 \% of the expected radius for granule $i$ at $t=t_{0,i}$
as long as $t\le t_{0,i}$ and otherwise zero. This region of avoidance
keeps new granules from  being formed in the immediate vicinity of
other newly formed granules. Therefore there are
two criteria a new generator point $(n)$ must fulfil when it is
generated, in order to be part of the tessellation,
\bean
\mathcal{W}(\vec{x}_n,t)& <& W_{thresh}\nonumber \\
\mathcal{A}(\vec{x}_n,t)&=&0\nonumber ~.
\eean

This ensures that there is room for a new granule and avoids
overpopulation of new granules later when the granules are growing. At
the same time the weight function evolves such that it makes new granules
appear only in intergranular lanes, because this is the location where
the weight function first drops below the threshold. 

In order to create the velocity-scale relation of the Sun, several
granular patterns with different scales must be superposed. This involves
creating a number of independent granulation patterns with different
typical scales as described above and adding the velocities. We have
chosen three layers, with typical granular radii of 1.3, 2.5, and 5.1
Mm respectively, with velocities and lifetimes following the observed
scaling relations. This choice corresponds to sizes ranging from those
of large granules 
to the size of mesogranules, and produces on average 880, 280 and 80
granules respectively, in the three size groups. The smallest scale is
set by the resolution and the largest scale is set by the box size
used in the simulation. 

Creating the velocity field in this way makes the velocity field
ordered on many scales, but also makes it too laminar. The vorticity
spectrum from the Voronoi driver is too weak compared to the simulations 
of \citet{Stein+Nordlund98}. This has been 
corrected by using a Helmholtz projection to separate the
rotational and irrotational part of the velocity field and then
amplifying the rotational part to the same
level as in the simulations by \citet{Stein+Nordlund98}. 

The velocity is quenched by a factor 
\begin{equation}
f_{quench}=\frac{1+\beta^{-2}}{1+3\beta^{-2}}\ ,
\end{equation}
depending on the plasma $\beta=P_{gas}/P_{Mag}$. In the
strongest magnetic field regions the quenching reduces the velocity
amplitude by $\sim 60$ \%, in order to reproduce the magnetic fields
ability to quench convective motions in the Sun.

The velocity -- scale
relation of the driver, averaged over 45 min, is shown in \Fig{fig:vspectrum}. 
To illustrate the time development of the driver, snapshots of corks
flowing in the time-dependent velocity field produce by the driver are
shown in \Fig{fig:corks}.

\subsection{Initial conditions}
Two initial stratifications were used. The first was taken from the FAL-C
model \citep{Fontenla+etal93} in the photosphere, chromosphere and
transition region, while in the corona a simple isothermal profile with
$T=10^6$ K was used (see Fig.\ref{fig:initstrat}). The second
was inspired by the work of \citeauthor{Carlsson+Stein02} \citep[see
for instance][]{Carlsson+Stein02}, who argue
for a highly time dependent, low mean temperature chromosphere with no
average temperature increase. This 
model has been criticised by \citet{Kalkofen01}, 
who argues for a warmer chromosphere. On the other hand observations of 
CO by \citet{Ayres02} point towards an even cooler chromosphere than the one
proposed by \citeauthor{Carlsson+Stein02}. 
The choice of chromospheric model turns out to have only a minor 
effect on the coronal dynamics and the coronal heating.

To keep the number of free parameters at a minimum we chose
to make the initial magnetic field a potential extrapolation of an
observed active region on the Sun. We chose to scale down a
high-resolution SOHO/MDI observation of active region 9114 from August
8, 2000. The cropped data spans $500\times 500$ pixels corresponding to
roughly $225\times 225$ Mm and was therefore scaled down to fit in the
computational box of $60\times 60$ Mm. The distribution of flux on the
solar surface seems to be approximately self-similar down to scales
$\sim$180 km in the network \citep{Schrijver+etal97b,Wang+etal95}, smaller 
than our resolution. If this also holds in active regions, as seems to
be the case \citep{Meunier03}, this would
justify our ``rescaling'' of the MDI magnetogram. It is however
necessary to point out that our computational box does not hold an
average size solar active region, but at most a small active region, and
therefore direct comparisons of these results with larger size
active regions should be done with care.

\ifnum\astroph=1
\FIGIV
\fi

\section{Results}
The simulation goes through a roughly five minute start--up period,
where neither thermal conduction nor radiative cooling are
turned on. This is related to the magnetic field initial condition. Since
the initial magnetic field is potential it is in a minimum energy state 
and is therefore initially not able to dissipate any energy.
In the first few minutes, thermal conduction and radiative cooling 
would therefore cool the corona monotonically, and make the atmosphere
tend to collapse. After the start--up period, when the magnetic field
has reached a dissipative state, the radiative cooling and the thermal
conduction is turned on. After another approximately five minutes
solar time a statistical equilibrium is reached, with a nearly
time independent temperature distribution. 

\subsection{Energy Balance}
The thermodynamic  equilibrium of the atmosphere is difficult to quantify
since the energy fluxes are very intermittent, but several conclusions
can still be drawn from a horizontal average of the energy flux
divergences, i.e.\ the deposited energy. The energy balance is shown in
\Fig{fig:balance}. This balance 
is for a typical point in time, and it is obvious that the heating in
the corona is decreasing with 
height, as found from TRACE data by a number of authors
\citep{Schrijver+etal99,Aschwanden+etal00b,Aschwanden+etal01}. The
detailed height dependence of the heating in individual loops is
investigated in a separate paper \citep[][]{Gudiksen+Nordlund03b}. 

It is noteworthy and interesting that the 
convective flux redistributes as much energy as the Spitzer conductivity
does just above the transition region. The heating provided by the
magnetic dissipation is transported 
by the Spitzer conductivity downwards from the transition region to
the upper chromosphere, where the Spitzer conductivity is no longer
effective due to the low temperature. The convective flux removes
energy at the same location because of evaporation of chromospheric
material into the corona. The energy loss in the lower corona and
transition zone due to the Spitzer conductivity is balanced by the
heating, which dominates the Spitzer conductivity at larger
heights. The optically thin radiative losses go to zero in the lower 
chromosphere and photosphere due to the quenched radiative cooling
function that we adopt.  The heating goes to very high values, but in spite
of that the temperature stays at chromospheric temperatures due to the
Newtonian cooling with which we represent the complex chromospheric 
radiation losses at these height. It is not possible to estimate whether
any further heating is needed in the chromosphere until we can
consistently treat both convective and radiative energy fluxes through
the lower boundary, as well as radiative losses in the optically thick lower
atmosphere. There should be no major effect on the 
conclusions drawn about the heating of the transition region and corona
due to this inadequacy, because we hold the temperature structure in this
region of the Sun close to what it is deduced to be observationally. 
Incorporating both enthalpy flux and an appropriate treatment of radiation
in a near ab-initio chromospheric model will take a considerable effort, 
and is a challenge for the future.

\ifnum\astroph=1
\FIGV
\fi

\ifnum\astroph=1
\FIGVI
\fi

\subsection{Magnetic field configuration and the dissipative heating}

The energy supplied to the corona originates from the Poynting
flux through the lower boundary.  In general, magnetic dissipative heating 
is generated from stressing of the magnetic field.  The stressing is largest 
in the photosphere and low chromosphere, where the plasma $\beta$ is larger 
than or of the order of unity in a significant fraction of the volume.
At these low heights the magnetic field still has 
a very intermittent structure, reflecting the intermittent distribution
set up by fluid motions in the photosphere and below (represented by
the initial and boundary conditions in our numerical model).  The plasma
$\beta$ is low in regions with high magnetic field strengths, but is still
high in intervening regions with low magnetic field strengths.

In the high chromosphere and above the magnetic field is nearly space-filling
and the plasma $\beta$ is less than unity almost everywhere (with the exception
of small neighbourhoods around magnetic null points).  The magnetic field thus
dominates the dynamics and tends to be `nearly force-free', in the sense that
the Lorentz force is small and the electric currents tend to be nearly
parallel with the magnetic field,
\begin{equation}\label{eq:alpha}
        \JJ \approx \alpha \BB .
\end{equation}
As is well known and easily proven, $\alpha$ must be constant along each
magnetic field line in a truly force-free field, but can certainly
vary from one field line to another. Also, since a dynamically
evolving coronal magnetic field cannot be entirely force-free, $\alpha$ should
not be expected to be exactly constant even along magnetic field lines.

In order to illustrate the variability of $\alpha$ over horizontal planes,
\Fig{fig:alpha} shows images of the quantity $\mid \vec{J} \cdot
\vec{B}\mid\ \approx \alpha B^2$ (we prefer to show this quantity rather than 
$\mid \vec{J} \cdot \vec{B}/ B^2\mid$, which would tend to be dominated
by regions with low $B$).    

If $\alpha$ was constant these images would effectively be the same as images 
of $B^2$.  The actual images show that $\alpha$ varies significantly
over horizontal 
planes, that it varies most rapidly at low heights, and that it certainly 
cannot be approximated with a constant at any height.  The larger horizontal
scale of the variability at larger heights is due to the fact that field lines 
(along which $\alpha$ is at least approximately constant) fan out as they 
reach higher, and hence spread the variability of $\alpha$ over larger patches.

We next comment on the absolute magnitude of $\alpha$, and its 
consequences for the overall field distribution.  Dimensionally, 
$\alpha$ is an inverse length; the length along a field line over which
it twists around itself.  As previously established by
\citet{Galsgaard+Nordlund96}, 
field lines that are stressed by a boundary typically are twisted only 
about once from end to end, corresponding to values of $\alpha$ of the
order of $1/L$, where $L$ is the field line length.  As illustrated by 
\Fig{fig:alpha}, $\alpha$ is only correlated over small patches near the lower
boundary.  This implies that the perturbations of the magnetic field line 
directions must be small (except in special regions such as in the
neighbourhood of quasi-separatrix layers \citep{Priest+Demoulin95}),
and hence that the overall 
perturbations of the magnetic field strength distribution, relative to
that of a potential field, must also be small. This means that to
obtain an overall picture of the magnetic field strength distribution,
a potential field extrapolation is sufficient in cases like this one, with no
large scale shear. 

\ifnum\astroph=1
\FIGVII
\fi

\ifnum\astroph=1
\FIGVIII
\fi

We expect the heating to generally be proportional to the electric
current squared and if the field is in a non-linear
force free state we also expect the electric current squared and thus the
heating to be proportional to the magnetic field strength
squared. \Figure{fig:current} shows a contour plot of the
distribution of heating with height in the atmosphere. One can see
that the heating is largest at low heights due to the stressing of
the magnetic field in a high $\beta$ environment. Through the
photosphere and chromosphere it declines by some
four orders of magnitude before reaching the transition region, 
above which the magnetic field is nearly force free and the
heating is indeed approximately proportional to the magnetic field
strength squared. Even though the electric current is highly
intermittent in horizontal planes, the current relative to the
magnetic field strength is almost constant along field lines above the
transition region, as is to be expected when $\alpha$ 
is constant along magnetic field lines. Heating with these characteristics
was proposed by \citet{Schrijver+etal99} and later deduced with data
from TRACE by \citet{Aschwanden+etal01}. Further studies of heating
models and observations have all supported these heating function
characteristics
\citep{Mandrini+etal00,Foley+etal02,Schrijver+Aschwanden02,Demoulin+etal03,Schmieder+etal03}.

The time variability of the heating depends critically on
height. \Figure{fig:heat-time} shows on which time scales the
energy dissipation varies, calculated as $k_t\, \sqrt{P_e}$ where $k_t$
is the time-wavenumber, and $P_e$ is the power spectrum of the
dissipated energy as a function of time at each point. This
produces a plot showing
on what timescales the energy is dissipated. The
energy dissipated is normalised to the total amount of dissipated
energy. Without this normalisation the curves would be widely
spaced along the y-axis, due to the exponential
decrease in total dissipated energy with height. In general most
of the energy is dissipated as short time scale events. In the 
upper chromosphere, the
energy comes primarily from events with a time
scale shorter than the time resolution of our snapshot data set
($\sim$ \mbox{38 s}). Slightly further up into the corona, the fraction
of energy dissipated at short time scales is
less pronounced. The rapid events most likely correspond to
``nano-flare'' like reconnection events.  With increasing spatial resolution,
we would expect to resolve smaller and smaller and hence increasingly
more short-lived events.

The energy supplied by the Poynting flux is distributed
initially between changes of kinetic, thermal, potential 
and magnetic energy, and dissipation. After the system has settled
into a statistical equilibrium the energy is only used to compensate for
magnetic and kinetic dissipation. The energy is dissipated in the 
whole atmosphere, with the main part in the photosphere and
chromosphere. The Poynting flux at the lower boundary is highly
intermittent but the total Poynting flux is close to
constant. Comparing the flux of energy through the lower boundary and
the heating rate in the corona gives an overall energy
conversion ``efficiency''. The time evolution of the efficiency is shown in
\Fig{fig:efficiency}(solid) during a 20 min period. During this time the
numerical resistivity was changed several times, apparently without
significant effect on the efficiency. The
efficiency is in the range 5--10 \%. That the 
changes in resistivity have little effect both on the Poynting flux into the
system and on the efficiency also confirms the results of
\citet{Hendrix+etal96} and \citet{Galsgaard+Nordlund96}. The radiative
losses of the corona is in the range 1--3 \% during the same time
period (\Fig{fig:efficiency}, dashed) making it less important than
the Spitzer conductivity, as expected.

The total amount of heating needed in active regions has been estimated
several times based on UV and X-ray fluxes, and typically gives
values in the range $10^6$--$10^7 \un{ergs\, cm^{-2}\, s^{-1}}$. These
values are in general hard to estimate precisely, since the total area of the
active region can be chosen based on different
criteria. \Figure{fig:totenergy} shows the 
average dissipated energy per second and square centimetre in our model,
in gas that is
emitting UV and X-rays ($T > 5\times 10^5 \un{K}$), in volumes below which the
photospheric magnetic field is higher than a certain threshold. If
only regions with high magnetic field strength below them are taken into
account the average energy dissipation can be as high as $8\times 10^6
\un{ergs\, cm^{-2}\, s^{-1}}$, while if one includes areas with lower 
field strengths the average dissipation rate declines, but remains
above $2\times 10^6 \un{ergs\,cm^{-2}\,s^{-1}}$ for the field
strengths shown in \Fig{fig:totenergy}.
These values of the average heating are well within the
observational limits. The total energy dissipated if the
chromosphere and transition region are also included is substantially larger
than these numbers, because most of the energy is dissipated in the high
field strength environment of the lower atmosphere.   

\ifnum\astroph=1
\FIGIX
\fi

\subsection{Stratification}
As limb observations by TRACE have indicated for some
time \citep[][ private communication]{Schrijver+McMullen99,Title+Schrijver03},
the transition region does not occur at a well defined height. 
The variation in transition region height, defined in our model as the height 
where the temperature 
rises above  $10^5 \un{K}$, is shown in \Fig{fig:transregion}
for one of the snapshots.  The height above the photosphere varies  
from 2.7 Mm to 12.3 Mm, with
an average of 5.0 Mm, and changes over small distances. The intermittency 
of the height of the transition region is a result of
the intermittent heating, which causes evaporation and re-condensation
of chromospheric material.

The stratification of the corona is intermittent on even the smallest
scales. The temperature is generally around $1 \un{MK}$ but there are
large deviations from this average temperature. There are differences 
in temperature of up to 0.7 MK over distances
comparable to TRACE's pixel size  ($\sim375 \un{km}$). The temperature 
gradients are most likely limited by the numerical resolution and are
therefore only lower limits. Because of the large temperature
gradients, modelling coronal loops as monolithic structures cannot be
a good approximation, thus making multi-thread
models of loops called for \citep{Aschwanden+etal00}. 

\ifnum\astroph=1
\FIGX
\fi

\Figure{fig:tstrat} shows that even at large heights in the corona 
there is gas at low temperatures. Evidence for cool gas suspended in
the corona has been found through limb observations with TRACE based on
opacity estimates \citep{Schrijver+McMullen99,Schrijver+etal99}. The cool
dense material in the lower part of the corona are primarily surges of
chromospheric material that flow into the corona, or condensations of coronal
material on its way down.  Higher up in the corona it is typically gas
undergoing catastrophic cooling \citep{Schrijver01}. At times surges
of material happen simultaneously in both foot points of a loop, which
brings large amounts of material to the top of the loop. 
If the loop is inclined relative to the vertical, 
accumulated mass at its top can force
the loop top down and make it almost horizontal along part of its
length, thus suspending the cold material in the corona. An example is the
large structure at roughly $(x,y)=(40,25)$ in
\Fig{fig:transregion}, which has a horizontal column density of 
up to $6\times 10^{20} \un{cm^{-2}}$ at a height of $8.0 \un{Mm}$,
making it a visible absorbing structure for UV at the limb. The
structure is long lived, having existed for 10 minutes at the end of the
simulation sequence. There are other similar structures, but too few 
to give a basis for firm estimates of typical characteristics.

The density has values that vary by at least 2 orders of
magnitude in the corona, over any horizontal slice. 
The average density and temperature in the corona depend primarily on
the amount of Poynting flux injected through the lower boundary
and on the choice of the chromospheric temperature structure, while
the horizontal variations of density and temperature primarily reflect 
the variations of properties from loop to loop. In general the highest
densities are in loop structures with low temperatures, persisting all
the way to the top of the atmosphere. A 2D histogram of the
Probability Density Function (PDF) of the density at each height 
for the FAL-C chromosphere is
displayed in \Fig{fig:rhostrat}, which shows that the horizontally 
averaged density stays
almost constant through the part of the corona that we model, with an average
number density close to $10^9 \un{cm^{-3}}$, but with values ranging
from $10^8$ to a few times $10^{10} \un{cm^{-3}}$, typical of or slightly
higher than observed for active regions loops \citep[see for
instance][]{Mason+etal99}.

\ifnum\astroph=1
\FIGXI
\fi

The velocities in the corona are not completely field aligned. In the
lower corona 80 \% of the velocity amplitude is along the magnetic
field, whereas higher in the corona the fraction drops to roughly 50
\%. In the 
lower corona the strong field regions have velocities that are almost
fully field aligned, while the lower field strength regions have
less prevalence for velocities along the magnetic field. 
In the upper corona the velocities are no longer strongly correlated 
with the magnetic field, and the velocities perpendicular to the
magnetic field are now as strong as those aligned with the magnetic field.
In general the velocity amplitudes in
the corona are highest just above the transition region, where the
impulsive energy releases can accelerate plasma to high
velocities. Velocities in this region reach as high as 400 km s$^{-1}$ 
but with an average of only a few tens of km s$^{-1}$.

\subsection{TRACE emission measure}\label{sec:trace}
Direct comparison of actual photon count rates between TRACE
observations and these simulations is complicated for several
reasons. \citet{DelZanna+Mason03} pointed out that the temperature
response functions of the TRACE filters are
based on the Chianti v2.0 database 
\citep{Landi+etal99}, which does not include recombinations to and
transitions for Fe VIII, and this affects the response functions
somewhat. Several ionisation equilibria models are published \citep[see for
instance][]{Mazzotta+etal98,Arnaud+Raymond92,Arnaud+Rothenflug85},
with differences that can give changes in the response function by up to a
factor of two \citep[][ private communication]{DelZanna+Mason03}. 
In this work, we have used the TRACE response function found by
\citet{DelZanna+Mason03}, with the ionisation states of
\citet{Mazzotta+etal98} and the elemental abundances by
\citet{Feldman92}, calculated at a constant electron pressure of
$10^{15} \un{K\, cm^{-2}}$. We have also made calculations using the
local electron pressure instead of a constant electron pressure, and
found that the difference is minor (typically less than 10 \%), but
the calculations are much more computationally expensive. 

\ifnum\astroph=1
\FIGXII
\fi

The images in \Fig{fig:TRACE} have been
produced by folding the temperatures and densities of the simulation with
the response function. In the 
run with the FAL-C chromosphere the calculated intensities in
the TRACE 171 and 195 filters are between 60-70 and 15-25 $\rm{DN\, s^{-1}\,
  pixel^{-1}}$. For individual loops chosen from the emulated TRACE
171 image, that is too high by almost a factor ten for the 171
filter and a factor 3 for the 195 filter compared to the values given by \citet{DelZanna+Mason03}. There are several
possible explanations for this discrepancy. The images show a forest
of loops of all heights. Some of the loops appear broad and diffuse
but this is often an effect of 
the perspective---most of these broad loops are not connected along the
projected axis. The numbers in \citet{DelZanna+Mason03} are for a
single, well isolated high reaching loop while the large number of loops in the
emulated images often are hard to isolate. The maximum coronal loop
height in the current model is limited to $\sim$ 30 Mm; loops
penetrating the upper boundary are not reliable, since the boundary
conditions cannot rigorously model the effects of the loop top on the
part of the loop included in the box. The count rates are therefore
hard to compare directly. The difference in the count
rates in the two filters indicates that there is lack of high
temperature material. Based on the previous work by
\citet{Galsgaard+Nordlund96}, we expect that higher resolution will create
more intermittent structures, creating a broader range of temperatures
and effectively shifting gas out of the 171 and 195 filters.

Cross sections and other properties for individual loops are examined
in a separate paper \citep{Gudiksen+Nordlund03b}.

\section{Conclusion}
The increase in computer performance and system memory in recent years
has made it possible to approach the solar coronal
heating problem from a new angle. Previously mainly qualitative models
have been developed, with assumptions of the magnetic field
structure and/or the atmospheric stratification that necessarily have been
simplified. This has kept conclusions about the underlying processes
indecisive. In recent years observational 
results coming from the SOHO and TRACE satellites have been able to put
constraints on the overall heating function, favouring a heating
function decreasing exponentially with height \citep{Schrijver+etal99,Aschwanden+etal00b,Aschwanden+etal01}.
Observations of nano- and micro-flares seem to indicate that even if one
extrapolates the distributions down to energies which are unobservable,
they are not able to supply all the
energy needed \citep{Aschwanden+Charb.02,Aschwanden+Parnell02,Parnell+Jubb00}.

\ifnum\astroph=1
\FIGXIII
\fi

\ifnum\astroph=1
\FIGXIV
\fi

By performing direct numerical simulations we have been able to
show that moving foot points of the magnetic field around in
a way consistent with the observed solar photospheric velocity fields
inevitably leads to an amount of energy dissipation in the corona that
is comfortably within the observational limits. The approach is nearly
{\em{ab initio}}, using only observed facts such as the average velocity
field properties, a realistic active region photospheric magnetic field,
a realistic optically thin cooling function, Spitzer conductivity, etc.

As anticipated by \citet{Parker72}, and as demonstrated numerically by
\citet{Galsgaard+Nordlund96} and \citet{Hendrix+etal96}, the magnetic
dissipation is expected to depend only weakly on the numerical resolution.
If anything it {\em increases} slightly as smaller scales are
resolved.  An aspect that {\em will} depend on numerical resolution,
however, is the time variability and intermittency 
of the heating.  As demonstrated by \citet{Galsgaard+Nordlund96}, the 
hierarchy of current sheets in which the magnetic dissipation takes place 
extends to ever smaller size as the numerical resolution is increased. 
It is presumably the intrinsic time variability from the reconnection 
events in this hierarchy that gives rise to the flare event size power
law distribution.  From this point of view, there is no conflict between 
the current model and the observations that have inspired the nano-flare 
coronal heating models.

The main results of the coronal modelling is that the total dissipated
energy is within the 
observational limits, making this heating mechanism at least a major
constituent (and an unavoidable one!) in heating the solar corona. 
The energy release should at least partially be observable as a 
distribution of flare-like events \citep{Galsgaard+Nordlund96},
but we cannot yet estimate the fraction of the energy released in the
form of nano--flares, or predict their energy distribution, since this
requires very high numerical resolution, outside the range of present
day computer capabilities. 

The magnetic field is in two distinctly different states
above and below the height where the $\beta=1$ layer is
located. Below, the field is controlled by the photospheric gas motions, 
and is in general in a non-relaxed state. The magnetic field there dissipates
energy at a high rate, since the magnetic energy density is high, the
magnetic field lines are short, and since
oppositely directed field lines may be forced
towards each other without the magnetic pressure having a significant
effect on the total pressure. 

In the present simulations roughly 90 \% of the
total dissipated energy is dissipated below the transition
region. Above the transition region the magnetic 
field assumes a non-linear near-force-free configuration, where the
dissipated energy is roughly proportional to the magnetic energy
density. Several observations point towards a heating function with
precisely these properties \citep{Mandrini+etal00,Foley+etal02,Schrijver+Aschwanden02,Demoulin+etal03,Schmieder+etal03}.

The results for the distribution of both temperature
and density are also well within the observed values, even though these
values might change slightly depending on the
stratification of the upper chromosphere. The temperature is
intermittent on the 
smallest scales, changing by as much as 0.7 MK on the scale
of a TRACE pixel size. The temperature has values from $10^4$ to
$3\times 10^6$ K in most of the corona, with a transition zone
defined as being at temperatures near $5\times 10^5$ K varying by as much as 9
Mm in height, similar to what has been inferred from TRACE observations
\citep[][ private communication]{Schrijver+McMullen99,Title+Schrijver03}. The
average density is almost constant with height in the corona, but has
values at each height spanning two orders of magnitude. 

From the point of view of the energy balance, the density and temperature
in the corona play different but tightly coupled roles.  On the one hand
the radiative cooling function changes relatively little with temperature, 
from a few times $10^5$ K to a few times $10^6$ K, while the cooling per 
unit volume is proportional to the square of the density.  On the other
hand heat conduction scales as $T^{7/2}$, but any increase in heat conduction
also results in more evaporation from the chromosphere, so tends to increase
the density in the corona.  So in this sense the temperature is not by
itself a sensitive probe for the heating in the corona, but has to be
taken together with density information, and with information about the
temperature gradients along the magnetic field.

Making a direct comparison with TRACE 
observations of an active region is hampered by both the question of
the statistical significance of a single observation, and by the
insufficient knowledge of the ionisation balance in the corona which makes
computed photon count rates uncertain by up to a factor of
two. Furthermore, the density in the corona is a function of the
stratification in the chromosphere, as commented upon earlier.

In spite of these difficulties the emission measures in the TRACE 171 and
195 {\AA} bands estimated from the simulations are not far from observed
values.  The discrepancies (mainly too much emission in the 171 band) is
most likely due to missing intermittency, in turn due to the limited 
numerical resolution.  As discussed in the previous Section, increased 
numerical resolution is expected to give a broader distribution of 
temperatures, reaching higher peak values.  This will reduce the emission
measure in the 171 band and will also affect the 195 band.

To reproduce the
hotter Yohkoh-loops will certainly
require a temperature distribution that extends to higher values than what
we find here. Increased heating will inevitably result if a more complete
representation of the solar velocity field is used.  In the current simulations
there is a resolution cut-off in the velocity driving towards smaller scales, 
and a box size cut-off
towards larger scales. It is interesting to note that scaling expressions
for coronal heating predict that the heating should scale with the product
of the velocity amplitude and the size scale of the motions, so there should
be significant additional heating from larger scales.  However, with the
turn-over time varying roughly as the square of the scale of motions, it
will take an ever increasing amount of computer capacity to recover these
contributions from larger scale motions.

A more accessible but less well defined contributor to 
additional heating would be active-region-specific velocity fields, such
as shearing motions, sunspot rotation, etc..  Such systematic motions
would cause systematically stressed magnetic fields, producing
locally enhanced heating rates in areas affected by the additional
systematic motions. Emerging flux is another contributor to heating
that is not treated in this simulation but happens continuously on the
Sun and will produce additional heating if included.

\acknowledgements{
BVG acknowledges support through an EC-TMR grant to the European
Solar Magnetometry Network.  The work of {\AA}N was supported in
part by the Danish Research Foundation, through its establishment
of the Theoretical Astrophysics Center. Computing time was provided 
by the Swedish
National Allocations Committee and by the Danish Center for Scientific
Computing. The authors thank G.\ Del Zanna for the improved
TRACE response function, R.\ Nightingale for general help with the
TRACE characteristics and V.\ Hansteen for providing the radiative
cooling function.
Both authors gratefully acknowledge the hospitality of LMSAL and ITP/UCSB
(through NSF grant No. PHY99-07949) during this work.
}

\bibliographystyle{aa}
\bibliography{ms}

\ifnum\astroph=0
\clearpage
\FIGI
\clearpage
\FIGII
\clearpage
\FIGIII
\clearpage
\FIGIV
\clearpage
\FIGV
\clearpage
\FIGVI
\clearpage
\FIGVII
\clearpage
\FIGVIII
\clearpage
\FIGIX
\clearpage
\FIGX
\clearpage
\FIGXI
\clearpage
\FIGXII
\clearpage
\FIGXIII
\clearpage
\FIGXIV
\fi
\end{document}